\documentclass[aps,a4paper,prstab,twocolumn,groupedaddress,nofootinbib]{revtex4-1}

\usepackage{graphicx}
\usepackage{mathtools}  
\usepackage{amssymb}
\usepackage{amsmath}
\usepackage{color}

\newtheorem{definition}{Definition}
\def\tsi{$\mathbf{x}^{\left(i\right)}$}
\def\tsj{$\mathbf{x}^{\left(j\right)}$}

\begin{document}


\title{Genetic Algorithm for Chromaticity Correction in Diffraction Limited Storage Rings}


\author{M. P. Ehrlichman}
\email[]{michael.ehrlichman@psi.ch}
\affiliation{Paul Scherrer Institut, Villigen, Switzerland}


\date{\today}

\begin{abstract}
A multi-objective genetic algorithm is developed for optimizing
nonlinearities in diffraction limited storage rings.  This algorithm determines
sextupole and octupole strengths for chromaticity correction that deliver optimized
dynamic aperture and beam lifetime.  The algorithm makes use of dominance constraints
to breed desirable properties into the early generations.  The momentum aperture
is optimized indirectly by constraining the chromatic tune footprint and
optimizing the off-energy dynamic aperture.  The result is an effective and computationally
efficient technique for correcting chromaticity in a storage ring while maintaining
optimal dynamic aperture and beam lifetime.
This framework was developed for the Swiss Light Source (SLS) upgrade project \cite{ipac15-sls2}.
\end{abstract}


\maketitle

\section{Introduction}
Multi-objective genetic algorithms have found much success
providing non-intuitive solutions to problems that are not adequately solved by analytic methods.
Such algorithms have been successfully applied to various aspects of accelerator design \cite{ga:hofler},
\cite{yves:thesis}, \cite{BorlandMOGA}, \cite{adelmann:arxiv}, and \cite{apisa:bazarov}.  
In this paper, a genetic algorithm is developed for optimizing dynamic aperture
and beam lifetime using sextupole and octupole strengths
in next-generation diffraction limited storage rings (DLSR).

The $1$-turn map for a DLSR contains strong nonlinearities.
Such machines use stronger focusing and lower dispersion to achieve an emittance that is below
the diffraction limit for some wavelengths of light.
The strong focusing of these machines leads to a large chromaticity, which must be corrected by placing strong
sextupoles in dispersive regions.  Furthermore, the low dispersion necessitates that the sextupole strengths be
even higher.  Such strong sextupole moments add strong nonlinearities to the lattice.  

These nonlinearities must be carefully designed to maintain adequate injection efficiency and
beam lifetime.  Nonlinearities induce tune shifts
and increase the sensitivity of particles to machine imperfections.
This reduces the dynamic aperture, which is the volume in 6D phase-space 
containing stable particle trajectories.  
Touschek scattering and residual gas scattering excite the phase space coordinates of particles stored in a ring.
A particle is lost if it exits the dynamic aperture.  This imposes a beam lifetime.
A reduced dynamic aperture also complicates injection by reducing the capture efficiency of
the machine.
The problem of nonlinear optics in a storage ring is to correct the chromaticity without 
inducing nonlinearities that degrade the other lattice properties too much.

The established technique for optimizing nonlinearities in a storage ring is resonant driving
term minimization and was developed
for the original SLS \cite{bengtsson-sls}.  It is a Lie algebra expansion
of the transfer map in driving terms that are functions of the sextupole strength.
These terms drive higher order chromatic tune shifts, amplitude dependent tune
shifts, and resonances.
A gradient optimizer is used to minimize a weighted vector of the driving terms.
The weights are determined by judging from the tune diagram and frequency maps
which resonances and tune shifts are most important.
This technique is straightforward to wield up to second order in sextupole strength.  Beyond 
second order, the number of driving terms which need to be minimized makes the
method less robust.  The complexity of the optimization space necessitates much trial and error
to locate good local minima, and the procedure becomes a bit of a dark art.

A storage ring is a complicated nonlinear oscillator.  Its transfer map is the concatenation
of many hundreds of linear and nonlinear maps.  Explicit algorithms yield only 
an incomplete control of particle motion in a storage ring.
This lack of analytic clarity makes accelerators good candidates for genetic algorithms, which typically
do not depend on knowledge of the system.  Genetic algorithms optimize a system
by selectively breeding trial solutions according to their fitness.

One example where a genetic algorithm performs better than perturbation techniques is in 
confining the chromatic and amplitude-dependent tune shifts.
Second order perturbation theory is unable to bend the chromatic and 
amplitude dependent tune shifts beyond second order.  The genetic algorithm 
developed here fits tune footprints into tighter areas using
higher orders of correction.

A well-designed genetic algorithm will encourage behaviors in the evolving population that, 
further down the road, lead to solutions with optimal objective functions.
For example, the algorithm presented here includes a dominance constraint (See Sec.~\ref{sec:domcom})
on the amplitude of the nonlinear dispersion.
A small-amplitude nonlinear dispersion is not one of the objective functions, but it is 
a property that a lattice with good objective functions will have.  
By applying a dominance constraint to the nonlinear dispersion, we are, in a sense, breeding characteristics
into the early generations, that in later generations will yield improved objective functions.

Genetic algorithms for the optimization of storage ring nonlinearities have been developed and
evaluated elsewhere \cite{BorlandMOGA}, \cite{ga:als}, \cite{brookhavenMOGA}, \cite{ga:gao}, and \cite{nim:MOPSO}.
The present application stands out in its use of dominance constraints to more efficiently evolve
the population.
The algorithm requires only modest computing resources.  On a Linux cluster consisting of $64$ E5-2670 Xeon
cores, it delivers solutions to $10$ variable problems in one or two days, and $20$ variable problems
in two or three days.
The scheme consistently does as well or better than Lie algebra approaches, 
and repeated attempts with different random seeds on different lattice variants suggest
that the solutions it finds are globally optimal.  
So the optimization scheme presented here allows for a lattice development cycle on the order of a couple days,  
and does not consume expensive computing resources.

Section \ref{sec:system} of this paper gives an overview of the system, listing the components out of which this optimization scheme
is built.  Section \ref{sec:moo} introduces multi-objective genetic algorithms, and 
Sec.~\ref{sec:optimization-problem} describes the optimization problem, including the physics behind the calculations of the objectives
and constraints.
In Sec.~\ref{sec:application} the optimization scheme is applied to upgrade prototypes of SLS 
and the proposed Armenian light source CANDLE \cite{candle-esls}.
Misalignment studies are conducted on the SLS upgrade lattice in Sec.~\ref{sec:ma}.

\section{System Architecture} \label{sec:system}
The optimizer is built within the {\tt PISA} framework \cite{pisa}, which specifies 
that the sorting algorithm be separated from the rest of the optimizer.
The sorting algorithm is implemented as a stand-alone binary that communicates with the
rest of the optimizer using a text-file based API.
This separation simplifies the coding and makes it trivial to switch between 
different sorting algorithms, such as {\tt SPEA2} \cite{spea2}
or {\tt NSGA2} \cite{nsga2}.

We use the {\tt aPISA} variant \cite{apisa:bazarov} of {\tt PISA}.  {\tt aPISA} modifies
the original {\tt PISA} framework by supporting dominance constraints.
{\tt aPISA} was originally developed for the design of the Cornell ERL injector.

Accelerator physics calculations are handled by calls to the {\tt Bmad} library \cite{bmad:2006}.
The top-level coding, which includes population management and breeding, parallelization, and additional physics
calculations, was developed at PSI and is coded in {\tt Fortran90}.  The parallelization paradigm is master-slave 
and is implemented using {\tt Coarrays}, which in Intel's Fortran compiler is implemented as a 
high-level language on top of {\tt MPI}.  

The cluster management software is Sun Grid Engine (SGE).  The cluster is composed of several $16$-core
E5-2670 compute nodes running $64$-bit Scientific Linux.  Typically $4$ of these nodes are used in
an optimization run.

\section{Multi-Objective Genetic Algorithms} \label{sec:moo}
The multi-objective optimization problem is formulated as \cite{deb}:
\begin{equation}
\left.
\begin{aligned}
\text{Minimize}\qquad   & f_m\left(\mathbf{x}\right),        && m=1,2,...,M;\\
\text{subject to}\qquad & g_j\left(\mathbf{x}\right)\geq 0,  && j=1,2,...,J;\\
                        & h_k\left(\mathbf{x}\right)=0,      && k=1,2,...,K;\\
                        & x_i^{\left(L\right)}\leq x_i \leq x_i^{\left(U\right)} && i=1,2,...,n.
\end{aligned}
\right\}
\end{equation}
$f_m$ are the objectives, which generally are competing.  $g_j$ are inequality constraints and
$h_k$ are equality constraints. $x^{\left(L\right)_i}$ and $x^{\left(U\right)}_i$ are upper and lower
bounds on variables.  A vector of variable strengths $\mathbf{x}=\left(x_1,x_2,...,x_n\right)^T$ 
is called an individual.

A genetic algorithm manages a population of individuals.
Every individual $i$ in the population is represented by a vector of variables \tsi.
For our purposes \tsi is real-valued, but in general it could contain integer, logical, and complex variables.
Each individual $i$ has an associated vector of objective values $\mathbf{f}^{\left(i\right)}$ and
constraint values $\mathbf{g}^{\left(i\right)}$ and $\mathbf{h}^{\left(i\right)}$.

The output of a multi-objective optimizer is of a different nature than that of a single-objective optimizer.
A single objective optimizer, or equivalently, an optimizer
which reduces $\mathbf{f}^{\left(i\right)}$ to a single value by weighting the individual objectives,
gives the user one particular $\mathbf{x}$ that is the best solution to the optimization problem it could find.  
A multi-objective optimizer, on the other hand, returns a population of $\mathbf{x}$'s.  
This returned population is an optimal surface in the objective space called a Pareto front.
For any individual on the Pareto front, no improvement to any one of its objectives can be 
achieved without worsening the others.  
The user of a multi-objective optimizer typically applies additional criteria when selecting a particular solution
from the Pareto front.

\subsection{Ranking the Population} \label{sec:ranking}
The ranking of individuals in a single objective optimization problem
is straightforward: the individual with the better objective value is preferred.  
Ranking in multi-objective optimization problem is more complicated.  The core concept is the
{\it dominance relationship}, which is a way of comparing any two individuals, say
\tsi and \tsj.
They are compared by asking the question: ``Does \tsi dominate \tsj?''
The dominance relationship is defined as \cite{deb}:
\begin{definition}\label{def:dom}
An individual \tsi is said to dominate another \tsj,
if both of the following conditions are true:
\begin{enumerate}
\item \tsi is no worse than \tsj in all objectives.
\item \tsi is strictly better than \tsj in at least one objective.
\end{enumerate}
\end{definition}

A sorting algorithm applies the dominance relationship to sort the population from best to worst.
There exist many different sorting algorithms.
Two algorithms that we have used are {\tt NSGA2} and {\tt SPEA2}.
We obtain similar results with both algorithms, but find that the populations
resulting from {\tt SPEA2} span the objective space more evenly.

For details on {\tt SPEA2} see Ref.~\cite{spea2}.  In short summary, dominance is determined
for every ordered pair of individuals in the population.  Each individual in the population is assigned
a strength which is the number of individuals it dominates.  Then, each individual is assigned
a fitness which is the sum of the strengths of all the individuals that dominate it.  A lower fitness
is better.  A `clumping' penalty is added to this fitness based on the shortest distance (in objective
space) of the individual to another individual.  This encourages the population to span a wider
region of the objective space.  Incidentally, the clumping penalty makes it unlikely for any two individuals
to have the exact same fitness, even if they are both not dominated by any other individual.

Individuals are ordered according to their fitness value.  
The lowest ranked individuals are deleted from the population.
This is typically the worst half or three-quarters of the population each generation.
The population is replenished by mating the surviving individuals.

Mating pairs are determined by drawing integers in a procedure called a {\it tournament}.  
Two or more random integers are drawn between $1$ and $N$, where
$N$ is the number of surviving individuals.  $1$ corresponds to the most fit, and $N$ to the
least fit.  The individual corresponding to the smallest of the drawn integers
is chosen for mating.  Thus, individuals with better fitness are more likely to reproduce.
Its mate is chosen through the same process.  This is repeated until enough pairs have been
selected to replenish the population.  Each pair generates two offspring.  

A two-step process generates two new children from each mating pair.
The first step is simulated binary cross-over \cite{deb:sbx}.  Say the two parents are
${\mathbf{x}^{\left(p1\right)}=\left\{x^{\left(p1\right)}_1,x^{\left(p1\right)}_2,...,x^{\left(p1\right)}_n\right\}}$ and
${\mathbf{x}^{\left(p2\right)}=\left\{x^{\left(p2\right)}_1,x^{\left(p2\right)}_2,...,x^{\left(p2\right)}_n\right\}}$.
They will produce two offspring, $\mathbf{x}^{\left(c1\right)}$ and $\mathbf{x}^{\left(c2\right)}$.

\begin{figure}
\includegraphics[width=\columnwidth]{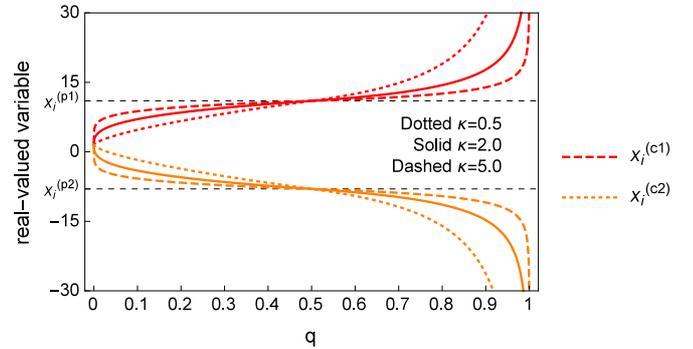}
\caption{The crossover operation takes the parent values $x_i^{\left(p1\right)}$ and $x_i^{\left(p2\right)}$ and a random
number $q$, and returns two offspring values $x_i^{\left(c1\right)}$ and $x_i^{\left(c2\right)}$.  Larger values of the 
parameter $\kappa$ result in `near-parent' offspring.  \label{fig:crossoverPlot}}
\end{figure}

Start with variable $x_1^{\left(\cdot\right)}$ and draw a random real $t$ between $0$ and $1$.  
Compare $t$ to $P_c$, which is a parameter between $0$ and $1$ that determines
how likely it is that a variable is simply copied from parent to child, as opposed to applying a stochastic function.
If $t > P_c$, then variable
$x^{\left(p1\right)}_1$ is simply copied to the child $x^{\left(c1\right)}_1$, and 
$x^{\left(p2\right)}_1$ copied to $x^{\left(c2\right)}_1$.
If $t \le P_c$, then draw another random real $q$ between $0$ and $1$.  Then,
\begin{equation}
\beta_q=\begin{dcases}
\left(2q\right)^\kappa & \text{if } q\leq 0.5\\
\left(\frac{0.5}{1-q}\right)^\kappa &\text{otherwise,}
\end{dcases}
\end{equation}
and
\begin{align}
x_1^{\left(c1\right)} &= \frac{1}{2}\left(\left(1+\beta_q\right)x_1^{\left(p1\right)} + \left(1-\beta_q\right)x_1^{\left(p2\right)}\right)\\
x_1^{\left(c2\right)} &= \frac{1}{2}\left(\left(1-\beta_q\right)x_1^{\left(p1\right)} + \left(1+\beta_q\right)x_1^{\left(p2\right)}\right),
\end{align}
where $\kappa$ is a parameter that controls the width of the distribution.  As depicted in Fig.~\ref{fig:crossoverPlot},
smaller values of $\kappa$ cause the offspring to explore a broader parameter space.  A typical value for $P_c$ is $0.8$.
This process is repeated for all $n$ variables.

Next a mutator is applied separately to $\mathbf{x}^{\left(c1\right)}$ and $\mathbf{x}^{\left(c2\right)}$.  For
each variable $i$ in $\mathbf{x}^{\left(c1\right)}$, draw a random real $t$ between $0$ and $1$.  Compare $t$ to $P_m$, a parameter
that determines how likely it is for a variable to undergo mutation.  If $t > P_m$, the variable is not mutated.  If $t \le P_m$, 
then draw a second random real $m$.  Variable $x_i^{\left(c1\right)}$ is adjusted according to
\begin{equation}
\beta_m=\begin{dcases}
\left(2m\right)^\kappa-1                & \text{if } m\le 0.5\\
1-\left(2\left(1-m\right)\right)^\kappa & \text{otherwise,}
\end{dcases}
\end{equation}
and
\begin{equation}
x_i^{\left(c1\right)} = x_i^{\left(c1\right)} + \sigma_\text{mut}\times \beta_m \text{.}
\end{equation}

\begin{figure}
\includegraphics[width=\columnwidth]{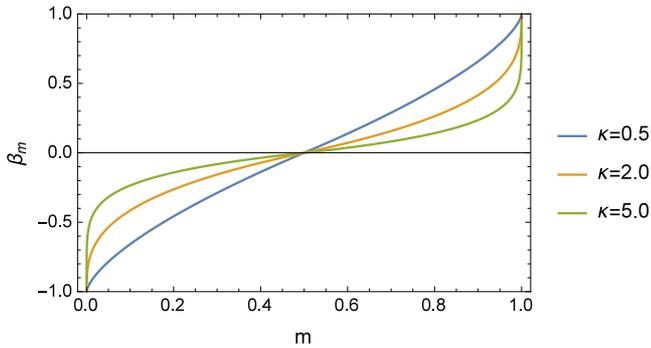}
\caption{The mutator operation takes a random real $m$ and adjusts an offspring variable by an amount $M\times \beta_m\left(m\right)$.
$M$ is a settable parameter which can be customized for each variable type.
Larger values of $\kappa$ make large mutations less likely.\label{fig:mutatorPlot}}
\end{figure}

A typical value for $P_m$ is $n^{-1}$, so that on average one variable is mutated per child.
$\sigma_\text{mut}$ is typically set to about $10\%$ of the reasonable variable strength.  $\beta_m$ is depicted in Fig.~\ref{fig:mutatorPlot}.

After $\mathbf{x}^{\left(c1\right)}$, the mutation process is repeated on each variable in 
$\mathbf{x}^{\left(c2\right)}$ and both are added to the population.  New individuals are generated until the population is fully replenished.  

\subsection{Progression of Generations}
Once the population is fully replenished, the worker processes calculate 
the objectives and constraints of the newly generated individuals.
The population, consisting of both the parents and children, is resorted and the least fit are deleted.  
Some of the children will be more fit and displace
the older individuals in the surviving population.  So is the overall fitness of the population improved.  
This process of evaluation, sorting, deletion, and replenishment
is looped continually.  Each loop is referred to as a generation.

As implemented, there is not actually a clear line between one generation and the next.  The algorithm is modified to improve
computational efficiency.  By far, the most time-consuming step is when the objectives and constraints of an individual
are calculated.  It could often be the case that many nodes in the cluster sit idle while
the last few individuals of a given generation are being evaluated.  To avoid this situation, the optimizer initially generates 
extra individuals equal to the number of cores in the cluster.  For example, if the population is set to $300$ and there are $64$ 
nodes in the cluster, the optimizer will initially generate $364$ individuals.   Whenever a core finishes evaluating an
individual, the individual is added to the population and the core is always immediately given a new unevaluated individual to process.  
Whenever the population reaches $300$, it is resorted, culled, and new individuals are bred.  These new individuals
replenish the pool of individuals awaiting evaluation.
This modification to the algorithm improves cluster efficiency.  The loss of a clear demarcation between the generations
does not seem to negatively impact the evolution of the population.

\subsection{Dominance Constraints}\label{sec:domcom}
Dominance constraints are a powerful type of constraint that is unique to multi-objective optimization.
It is implemented by modifying the dominance relationship.
We apply the dominance relationship as implemented in {\tt aPISA} \cite{apisa:bazarov}.

In addition to calculating objective values $\mathbf{f}\left(\mathbf{x}^{\left(i\right)}\right)$ for each individual $i$, we also calculate a vector of 
constraint values $\mathbf{c}\left(\mathbf{x}^{\left(i\right)}\right)$.  For example, say $c_1$ is a dominance constraint for the off-momentum horizontal closed 
orbit $x_\text{inj}$ at the injection point, and we wish to constrain $\left|x_\text{inj}\right|$ to be 
less than $x_\text{max}$.  Then, $c_1 = x_\text{max}-\left|x_\text{inj}\right|$.
If $c_1$ is negative, it indicates that the constraint
is violated.  The magnitude of $c_1$ represents the degree to which it is violated.

Dominance constraints are implemented by replacing the 
ordinary definition for dominance in Def.~\ref{def:dom} with Def.~\ref{def:domcom}.
An individual is called {\it infeasible} if it violates
any of its dominance constraints, else it is called {\it feasible}.
\begin{definition}\label{def:domcom}
An individual \tsi is said to constraint-dominate an individual \tsj
if any of the following conditions are true:
\begin{enumerate}
\item \tsi is feasible and \tsj is not.
\item \tsi and \tsj are both feasible, and \tsi dominates \tsj as in Def.~\ref{def:dom}.
\item \tsi and \tsj are both infeasible, and both
\begin{enumerate}
\item \tsi is no worse than \tsj in all constraints.
\item \tsi is strictly better than \tsj in at least one constraint.
\end{enumerate}
\end{enumerate}
\end{definition}

The behavior of a genetic algorithm implementing dominance constraints flows through three phases:
\begin{enumerate}
\item Random population, possibly containing no feasible individuals, spans variable space, sorted according to severity of constraint violations.
\item Entire population is feasible and spans feasible region of variable space, sorted according to objective values.
\item Population spans variable space that approaches the Pareto optimal objective space.
\end{enumerate}

Notice that if an individual violates any of its dominance constraints, then its objective values are not taken into account
during ranking.  Therefore, to save computing time,
objective values are calculated only for individuals which do not violate any dominance constraints.  Dominance constraints are based
on variable bounds, closed orbit amplitudes, and chromatic tune shifts.  These quantities are orders of magnitude quicker to evaluate
than objective values, which are based on particle tracking.

\section{Evaluation of Individuals}\label{sec:optimization-problem}
The design problem is to correct the chromaticity while maintaining acceptable 
dynamic aperture and beam lifetime.  
The beam lifetime would be maximized by
maximizing the momentum aperture, which is the largest momentum kick that an initially on-axis particle
can receive without being lost downstream.  
The momentum aperture can vary throughout the lattice and is typically calculated
element-by-element or in fixed steps.  
It is computationally expensive to calculate, so instead we optimize the off-energy dynamic aperture
and apply a dominance constraint to the chromatic tune footprint.  
The results in Sec.~\ref{sec:application} show that this is an effective proxy for the momentum aperture.  

\subsection{Objectives}
\begin{figure}
\centering
\includegraphics[angle=-90,width=\columnwidth]{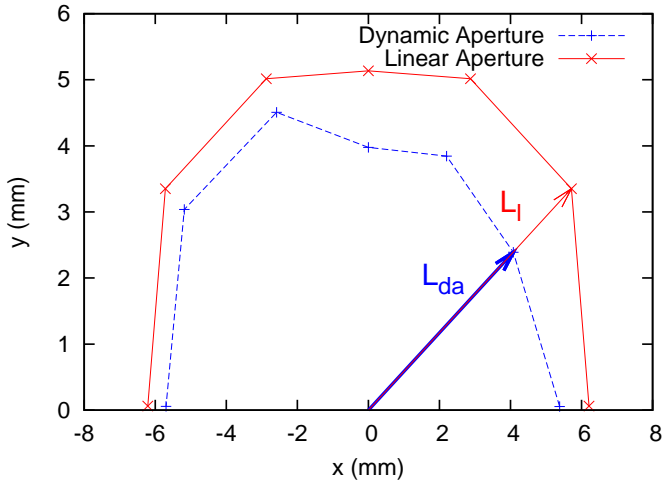}
\caption{Dynamic aperture is calculated using an element-by-element tracking code.
Particles are tracked for $200$ turns.  The aperture is found using a binary
search for particle loss.  In this plot $N_\text{angle}=7$.\label{fig:da}}
\end{figure}

The dynamic apertures are calculated relative to the linear aperture.  The linear aperture is
the smallest aperture found by
projecting the beam chamber from each point in the lattice to the injection point
using the linearization of the map about the particle momentum.  
The linear aperture, in general, depends on the particle momentum.  
The on-energy linear aperture does not depend on the sextupole and multipole strengths, 
but the off-energy linear aperture does.  The objective function
is formulated relative to the linear aperture such that $1.0$ is perfectly 
bad and $0.0$ is perfectly good,
\begin{multline}
f\left(\mathbf{x}\right)=\\ 
\frac{1}{N_\text{angle}}\sum\limits_{N_\text{angle}}
\begin{dcases}
\left(\frac{L_\text{l}\left(\mathbf{x}\right)-L_\text{da}\left(\mathbf{x}\right)}
{L_\text{l}\left(\mathbf{x}\right)}\right)^2, & \text{if } L_\text{da}<L_\text{l}\\
0, & \text{otherwise,}
\end{dcases}
\label{eqn:obj}
\end{multline}
where $N_\text{angle}$ is the number of rays along which the aperture is calculated.  $L_\text{l}$
is the length of the linear aperture ray.  $L_\text{da}$ is the length of the dynamic aperture ray.
$L_\text{l}$ and $L_\text{da}$ are depicted in Fig.~\ref{fig:da}.

The conditional in Eq.~\ref{eqn:obj} reflects our design philosophy that a machine with 
optimized nonlinearities should behave as if it were linear.  The objective function is 
not rewarded when the dynamic aperture exceeds the linear aperture.

Three objectives are used in our multi-objective optimization problem:
\begin{align}
f_0\left(\mathbf{x}\right)&=\text{on energy dynamic aperture}\\
f_+\left(\mathbf{x}\right)&=\text{dynamic aperture at $\Delta E_\textrm{+DA}$}\\
f_-\left(\mathbf{x}\right)&=\text{dynamic aperture at $\Delta E_\textrm{-DA}$},
\end{align}
where $\Delta E_\textrm{+DA}$ and $\Delta E_\textrm{-DA}$ specify the energy offset where dynamic aperture is evaluated.
Typical energy offsets are $2\%$ or $3\%$.

\subsection{Constraints}
Three constraint techniques are used: dominance constraints, modified objective functions,
and variable space projection.

Five constraints are implemented as dominance constraints.  They are
\begin{enumerate}
\item $c_\textrm{mag}$: Boundary on sextupole and multipole strength.
\item $c_{+co}$: Global bound on nonlinear dispersion at $\Delta E_\textrm{+DA}$.
\item $c_{-co}$: Global bound on nonlinear dispersion at $\Delta E_\textrm{-DA}$.
\item $c_{+\chi}$: Confine chromatic tune footprint between $0$ and $\Delta E_{+\chi}$.
\item $c_{-\chi}$: Confine chromatic tune footprint between $0$ and $\Delta E_{-\chi}$.
\end{enumerate}

The constraint on sextupole and multipole strength is calculated as,
\begin{equation}
c_\textrm{mag}=\sum\limits_i^\textrm{magnets}
\begin{dcases}
\frac{x_i^{\left(U\right)}-K_i}{\left|x^{\left(U\right)}\right|+\left|x^{\left(L\right)}\right|}, & K_i > x_i^{\left(U\right)} \\
\frac{K_i-x_i^{\left(L\right)}}{\left|x^{\left(U\right)}\right|+\left|x^{\left(L\right)}\right|}, & K_i < x_i^{\left(L\right)} \\
0, & \textrm{otherwise,}
\end{dcases}
\end{equation}
where $K_i$ is the strength of magnet $i$, and $x_i^{\left(U\right)}$ and $x_i^{\left(L\right)}$ are the upper
and lower bounds on the magnet strength.

The two constraints on global nonlinear dispersion are calculated as,
\begin{multline}
c_{\pm co} =\\
\frac{1}{x^{\left(\textrm{co}\right)}}\left(x^{\left(\textrm{co}\right)}
-\max\limits_{i\in\textrm{all elements}}\textrm{abs}\left(x_{\textrm{co},i}-\eta_i\times\Delta E_{\pm\textrm{DA}}\right)\right),
\end{multline}
where $x^{\left(\textrm{co}\right)}$ is the maximum allowed closed orbit, usually set to millimeter or so.  
$x_{\textrm{co},i}$ and $\eta_i$ are the closed orbit and ordinary dispersion at element $i$. 

The positive chromatic footprint between $0$ and $\Delta E_{+\chi}$ is constrained
to cross neither the half-integer nor the integer resonances.  It is calculated by dividing the interval from $\Delta E=0$
to $\Delta E=\Delta E_{+\chi}$ into $N_\chi$ equal segments.  
The horizontal and vertical tunes are calculated by linearizing the optics 
about each $\Delta E_{j} \in \left\{\Delta E_1,\Delta E_2,...,\Delta E_{+\chi}\right\}$.
The smallest of these energy offsets that results in an unstable transfer matrix, or a horizontal or vertical tune
that crosses an integer or  half-integer resonance, is used to calculate the value of the dominance constraint.  The value 
is calculated as,
\begin{equation}
c_{+\chi} = -1+\frac{j-1}{N_\chi}.
\end{equation}
If all $\Delta E_{+j}$ are stable and do not cross the half-integer or integer, $c_{+\chi}$ is set to $0$.
A similar procedure is used to constrain the negative chromatic tune footprint $c_{-\chi}$.  

Two constraints are implemented by modifying the objective functions.
The first is a constraint on the minimum size of the off-energy linear aperture and it modifies the off-energy objective
functions $f_+$ and $f_-$.  
The constraint prevents a failure condition where the optimizer improves the off-energy objectives by making the linear
aperture tiny, rather than by growing the dynamic aperture.  Setting this constraint to $2$ or $3$
mm is usually sufficient to avoid the condition.  If the shortest linear aperture ray 
at $\Delta E_\textrm{+DA}$ or $\Delta E_\textrm{-DA}$ is shorter than this constraint, 
then $f_+$ or $f_-$ is set to a perfectly bad value of $1$.

The second constraint implemented by modifying the objective functions modifies the on-energy objective function $f_0$.  
It confines the on-energy amplitude dependent tune shift (ADTS).  
Along the two DA rays closest to the horizontal axis, the horizontal and vertical tunes are calculated.
Along the vertical ray, the vertical tune is calculated.  This is because the horizontal motion for
large vertical and small horizontal offsets is dominated by coupling,
rather than by the horizontal optics.
The tunes are calculated by summing the element-by-element phase advance
in normal mode coordinates.  If the tunes of the 
particle cross the half-integer or integer resonances, then the tracking code considers the particle lost.  
The apertures along the two horizontal rays and one vertical ray define 
a `clipping' box.  When $f_0$ is calculated, all DA rays are clipped at the box.

The chromaticity correction is applied by projecting the $N_\text{cs}$-dimensional space of
chromatic sextupole strengths onto the $\left(N_\text{cs}-2\right)$-dimensional surface
on which the horizontal and vertical chromaticities have the desired values.
This is possible because chromaticity depends linearly on the sextupole strength.
$N_\text{cs}$ is the number of chromatic sextupole families in the lattice.
A chromaticity response matrix $\mathbf{A}$ is determined numerically,
\begin{equation}
\mathbf{A}=\begin{pmatrix}
\frac{d\chi_x}{dK_1} & \frac{d\chi_x}{dK_2} & \cdots & \frac{d\chi_x}{dK_\text{Ncs}}\\
\frac{d\chi_y}{dK_1} & \frac{d\chi_y}{dK_2} & \cdots & \frac{d\chi_y}{dK_\text{Ncs}}
\end{pmatrix}.
\end{equation}
The Moore-Penrose pseudoinverse $\mathbf{A}_p$ of $\mathbf{A}$ is calculated via 
singular value decomposition (SVD) \cite{matrix-comp}.
Then the thin-QR decomposition \cite{matrix-comp} $\mathbf{Q}_1$ of $\mathbf{I}-\mathbf{A}_p\mathbf{A}$ is taken, 
where $\mathbf{I}$ is the $N_{cs}\times N_{cs}$ identity matrix.  
Note that $\mathbf{Q}_1 \in \mathbb{R}^{N_{cs}\times N_{cs}-2}$.  

With $\mathbf{A}_p$ and $\mathbf{Q}_1$ in hand, take any vector $\vec{\omega} \in \mathbb{R}^{Ncs-2}$.
The chromatic sextupole strengths $\vec{K}$ given by
\begin{equation}
\vec{K}=\mathbf{A}_p\begin{pmatrix}\chi_{x0}\\\chi_{y0}\end{pmatrix}+\mathbf{Q}_1\vec{\omega}
\end{equation}
result in chromaticities of $\chi_{x0}$ and $\chi_{y0}$.

The algorithm does not operate directly on the chromatic sextupole strengths.  Instead it operates
on $\vec{\omega}$, thus constraining the chromaticities to the desired values and reducing
the dimension of the variable space by $2$.

\section{Application}\label{sec:application}
Here the genetic algorithm is applied to a prototype lattice for the SLS upgrade, and also the proposed Armenian light source CANDLE.

The SLS upgrade is a $2.4$ GeV storage ring built of $12$ arcs which consist of $5$ longitudinal
gradient bends (LGB) plus $2$ half-bend longitudinal gradient dispersion suppressors.  There are $3$ types of straight (short, medium, and long)
which reduce the periodicity to $3$.  The lattice uses anti-bends to focus the
dispersion into the LGBs to minimize the radiation integral ${\cal I}_5$ \cite{lgb-ls}.
The lattice parameters are summarized in Tab.~\ref{tab:lats}.
\begin{table}[h]
\begin{tabular*}{1.00\columnwidth}{@{\extracolsep{\fill}}lcc}
\hline
\hline
      & SLS upgrade \cite{ipac15-sls2} & CANDLE \cite{candle}\\
\hline
Circumference (m)                                  & $287.25$            & $258$ \\
Emittance $\left(\textrm{pm}\right)$                & $137$               & $1091$ \\
Periodicity                                         & $3$                 & $16$ \\
Topology                                            & $12\times 7$BA      & $16\times $4BA\\
$Q_x$                                               & $37.383$            & $24.700$ \\
$Q_y$                                               & $10.280$            & $14.368$ \\
Nat. chrom. $\chi_x$                                & $-64.9$             & $-40.6$ \\
Nat. chrom. $\chi_y$                                & $-34.5$             & $-26.5$ \\
Peak dispersion $\left(\textrm{cm}\right)$          & $4.9$               & $9.3$ \\
\# chromatic sextupole fam.                         & $4$                 & $8$ \\
\# harmonic sextupole fam.                          & $9$                 & $0$ \\
\# octupole fam.                                    & $10$                & $0$ \\
\hline
\hline
\end{tabular*}
\caption{SLS upgrade is a longitudinal gradient bend plus anti-bend based replacement for the SLS storage ring.  
CANDLE is a proposed Armenian light source based on combined function magnets.
\label{tab:lats}}
\end{table}

For the SLS upgrade lattice, the objective functions are the on-energy dynamic aperture and the
dynamic aperture at $-3\%$ and $+3\%$.  The chromatic footprint between $-5\%$ and $+5\%$
is constrained such that $37.0<Q_x<37.5$ and $10.0<Q_x<10.5$.  The amplitude-dependent tune shift
as described in Sec.~\ref{sec:optimization-problem} is constrained to this same region.

For CANDLE, the objective functions are the on-energy dynamic aperture and the
dynamic aperture at $-2\%$ and $+2\%$.  The chromatic footprint between $-3\%$ and $+3\%$
is constrained such that $24.5<Q_x<25.0$ and $14.0<Q_x<14.5$.  The amplitude-dependent tune shift
as described in Sec.~\ref{sec:optimization-problem} is constrained to these same regions.

The optimization parameters for both lattices are summarized in Tab.~\ref{tab:go}.

\begin{table}[h]
\begin{tabular*}{1.00\columnwidth}{@{\extracolsep{\fill}}lcc}
\hline
\hline
      & SLS upgrade & CANDLE\\
\hline
\# nonlin. mag. fam.                                        & $23$              & $8$ \\
\# variables                                                & $21$              & $6$ \\
$c_\textrm{mag,sext.}$ $\left(K_\textrm{sext}\right)^{*}$   & $500.0$           & $500.0$ \\
$c_\textrm{mag,oct.}$ $\left(K_\textrm{oct}L\right)^{*}$    & $500.0$           & $500.0$ \\
$\Delta E_\textrm{+DA}$                                     & $3\%$             & $2\%$ \\
$\Delta E_\textrm{-DA}$                                     & $-3\%$            & $-2\%$ \\
$\Delta E_{+\chi}$                                          & $5\%$             & $3\%$ \\
$\Delta E_{-\chi}$                                          & $-5\%$            & $-3\%$ \\
Footprint$^{**}$ $Q_\textrm{x,min}$, $Q_\textrm{x,max}$     & $37.0$, $37.5$    & $24.5$, $25.0$ \\
Footprint$^{**}$ $Q_\textrm{y,min}$, $Q_\textrm{y,max}$     & $10.0$, $10.5$    & $14.0$, $14.5$ \\
\hline
\hline
\end{tabular*}
\caption{Parameters for Genetic Optimizer.  $^*$The sextupole and octupole quantities have been
normalized by $n!$. $^{**}$The footprint constraints apply to both chromatic tune shift and on-energy ADTS.
\label{tab:go}}
\end{table}

\subsection{SLS Upgrade Lattice}

\subsubsection{Layout and Linear Lattice Considerations}
The following were taken into consideration during the design of the layout and linear optics
of the SLS upgrade lattice in order to improve the nonlinearities.

\begin{enumerate}
\item The ADTS is suppressed if the horizontal tune is close to
\begin{equation}
Q_\textrm{opt}=\frac{2q+1}{2}N,
\end{equation}
where $N$ is the periodicity of the machine and $q$ is some integer \cite[Chapter~14.3.1]{wiedemann}.  The periodicity
of SLS is $3$.  Thus ADTS is reduced by selecting a horizontal tune near one of
the following: $\left\{...,\,34.5,\,37.5,\,40.5,...\right\}$.
\item Chromatic sextupoles are placed where dispersion is large, and where either $\beta_x<<\beta_y$ or $\beta_x>>\beta_y$.  Harmonic 
sextupoles are placed in dispersion-free regions where $\beta_x>>\beta_y$, $\beta_x<<\beta_y$, or $\beta_x\approx\beta_y$.
\item The arcs are constructed from five identical unit cells.  The horizontal and vertical
phase advances per unit cell are $0.4$ and $0.1$ radians, respectively.  Over the five unit cells, the
lowest order chromatic and geometric resonant driving terms are canceled out \cite{bengtsson-sls}.
\end{enumerate}

\subsubsection{Optimization}
The population size for the SLS upgrade optimization is $300$ and begins with a pool of $364$ unevaluated, randomly generated, individuals.
Each individual is described by $21$ variables representing the
$23$ magnet strengths.  The strengths are bounded by $c_\textrm{mag,sext.}$ and $c_\textrm{mag,oct.}$, as given in Tab.~\ref{tab:go}.
Each seed is farmed via MPI to a CPU which evaluates its constraint and objective values.  As soon as $300$ individuals have been evaluated,
the population is sorted and the $150$ least fit are deleted.  A four-way tournament is used to select parent pairs from
the surviving population.  From each pair, simulated binary crossover plus mutation is used to generate two new children.  The new children
are added to the pool of unevaluated individuals, and the process repeats.

The initial random population contains no individuals which satisfy the dominance constraints (i.e.~all individuals are infeasible).  
Therefore the population members initially compete for who has the least-bad constraint violations.  For the particular 
optimizer run shown here,
the first feasible individual appears at generation $21$, and at generation $41$ the population consists entirely of individuals 
which satisfy all of the dominance
constraints.  From generation $1$ to $20$ requires $2$ minutes, and from generation $20$ to $41$ requires $14$ minutes.  This
first stage of the optimization proceeds quickly because infeasible individuals are evaluated only for off-momentum closed orbit amplitudes
and tune footprints.  During the remainder of the optimization run, individuals are competing based on dynamic aperture.
The entire optimization completes after $45$ hours to converge at generation $616$.  Computing resources are $64$ E5-2670 Xeon cores.
These compute time requirements are generally indicative of the resources required by the algorithm on the SLS upgrade.

Following the optimization run, the seeds with the most promising objective
function values are selected by hand for further evaluation.  Further evaluation includes on- and off-energy
rastered survival plots, higher resolution chromatic and ADTS tune footprints, momentum aperture, and Touschek lifetime 
evaluation.
Using these additional evaluations, the lattice designer selects a best individual.

\begin{figure}
\includegraphics[angle=-90,width=\columnwidth]{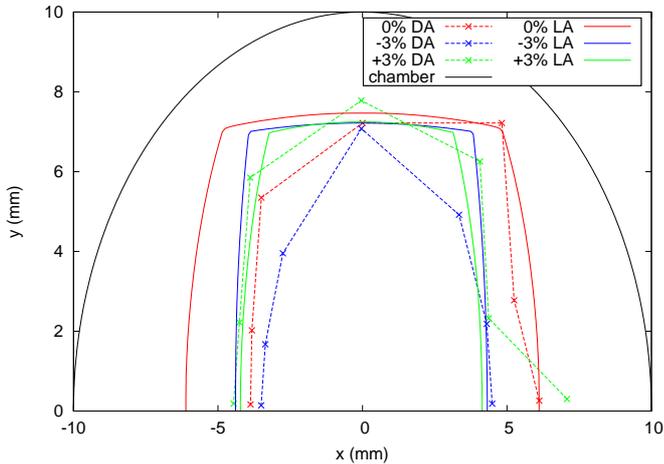}
\caption{Dynamic aperture for SLS upgrade at $0\%$, $-3\%$, and $3\%$ energy offset.
Dashed lines are dynamic aperture, solid lines are linear aperture.  
At the injection point $\beta_x=3.3$ m and $\beta_y=6.5$ m.
The maxima throughout the machine are $\beta_x=8.7$ m and $\beta_y=11.7$ m.
\label{fig:dc01a-da}}
\end{figure}

\begin{figure}
\includegraphics[angle=-90,width=\columnwidth]{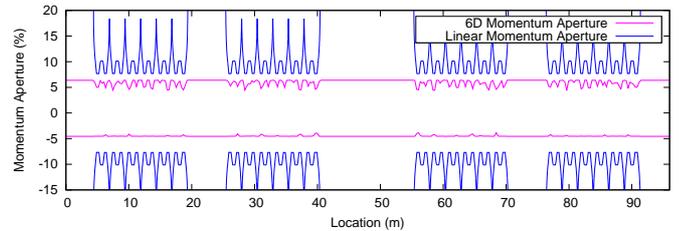}
\caption{Momentum aperture of SLS upgrade.  Linear aperture is calculated using an element-by-element
linearization of the optics.
The 6D aperture is calculated with full 6D tracking including radiation damping and synchrotron oscillation for $2.5$ periods
of the linear synchrotron tune.
The asymmetry of the 6D momentum aperture is the influence of nonlinear momentum compaction.  The RF voltage is set such
that the linear calculation gives a $\pm5\%$ bucket.
\label{fig:dc01a-ma}}
\end{figure}

The dynamic aperture of this individual at $0\%$, $-3\%$, and $+3\%$ is shown in Fig.~\ref{fig:dc01a-da}.  Figure \ref{fig:dc01a-ma}
compares the momentum aperture and linear momentum aperture.  The linear momentum aperture is calculated by linearizing
the one-turn map at each element.  The reference Touschek lifetime is calculated from the linear RF bucket height, 
and is taken as the benchmark against which to judge the effectiveness of the algorithm.

The assumptions used for the Touschek lifetime calculation are shown in Tab.~\ref{tab:calc-tl}.
The lifetime exceeds the reference lifetime
because the nonlinearity of the longitudinal phase space causes it to exceed the dimensions of the linear RF bucket,
and the momentum aperture is not otherwise limited by the transverse nonlinearities.

Recall that the genetic algorithm does not directly optimize the Touschek lifetime nor momentum aperture.  Rather,
it constraints the chromatic and amplitude-dependent tune footprints and maximizes the dynamic aperture
area at $0\%$ and $\pm3\%$.
The element-by-element variation in the momentum aperture is small.  This indicates that the Touschek lifetime
is limited by the longitudinal dynamics, and not by nonlinearities in the transverse optics.
Judging by this result, an off-momentum dynamic aperture optimization plus tune footprint constraint is a
valid proxy for optimizing the Touschek lifetime and momentum aperture.

\begin{table}[h]
\begin{tabular*}{1.00\columnwidth}{@{\extracolsep{\fill}}lcc}
\hline
\hline
         & SLS upgrade & CANDLE \\
\hline
Horiz. Emittance (pm)                                     & $137.$      & $1091.$ \\
Vert. Emittance (pm)                                      & $10.0$      & $10.0$ \\
Current per Bunch (mA)                                    & $1.0$       & $1.0$  \\
Number of particles per bunch ($10^9$)                    & $6.0$       & $5.4$  \\
Bunch length (mm)                                         & $0.261$     & $0.494$ \\
6D Touschek lifetime (hr)                                 & $4.58$      & $3.82$ \\
Reference Touschek lifetime (hr)                          & $4.35$      & $3.63$ \\
\hline
\hline
\end{tabular*}
\caption{Touschek lifetime calculation and assumptions.  Bunch length is natural without 3rd harmonic cavity.
The currents for both SLS upgrade and CANDLE assume a $500$ MHz RF system.  
The horizontal emittance is that given by the radiation integrals calculation.
The Touschek lifetime is calculated with 6D tracking including radiation damping and synchrotron oscillations.
The reference Touschek lifetime is calculated for the SLS upgrade by assuming a $5\%$ momentum acceptance everywhere, and 
by assuming $3\%$ for CANDLE.
\label{tab:calc-tl}}
\end{table}

\begin{figure}
\includegraphics[angle=-90,width=0.33\columnwidth]{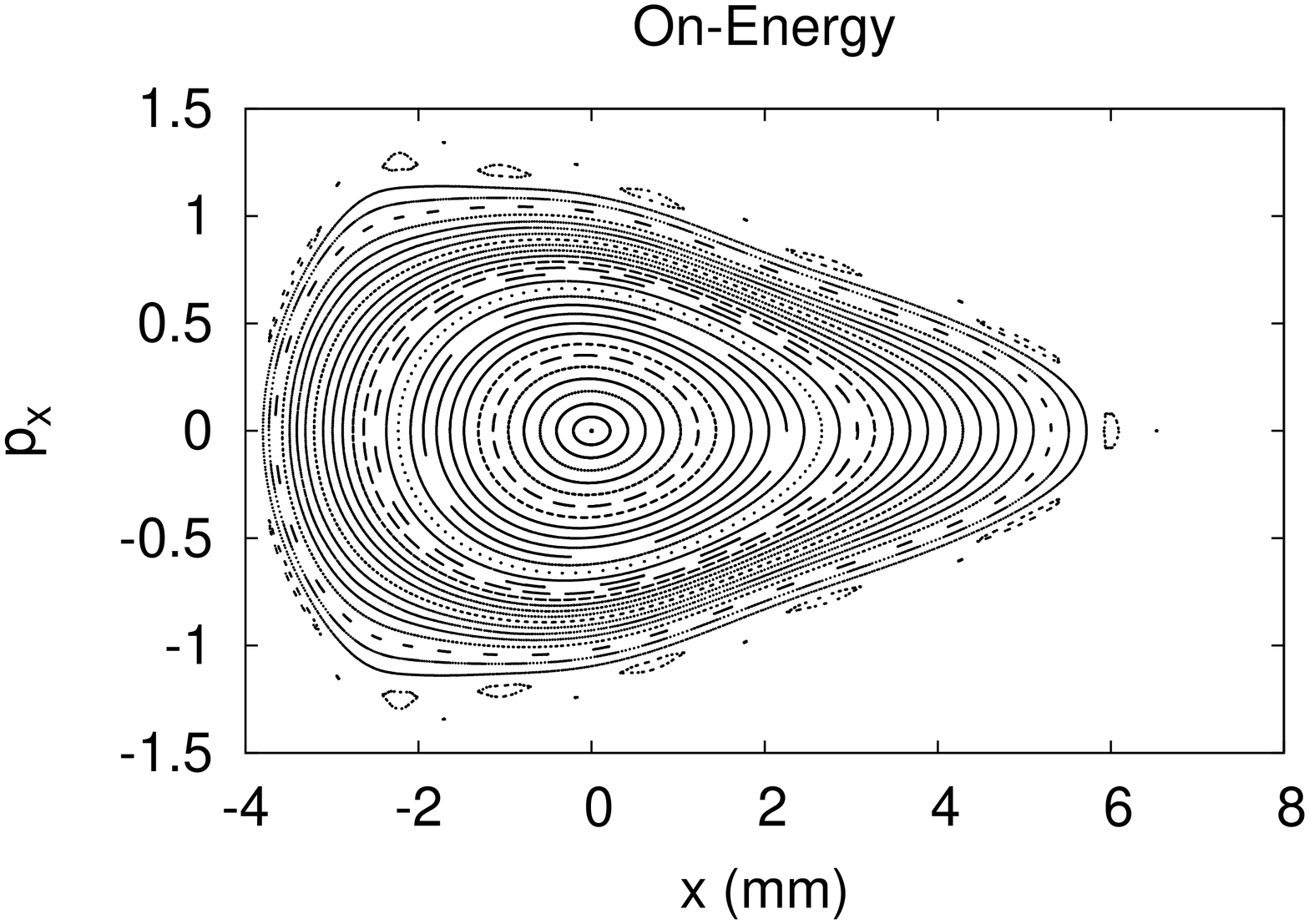}\hfill
\includegraphics[angle=-90,width=0.33\columnwidth]{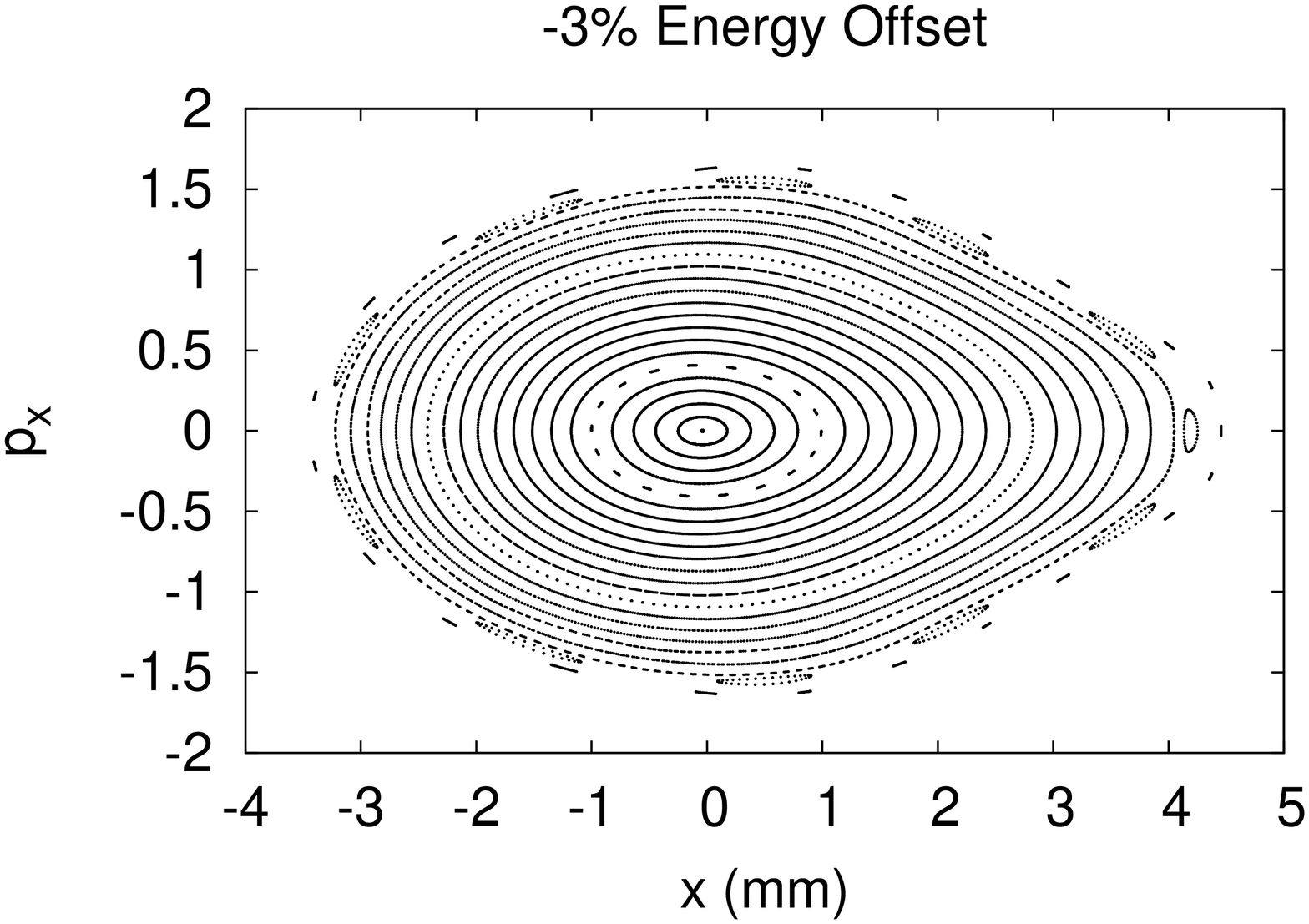}\hfill
\includegraphics[angle=-90,width=0.33\columnwidth]{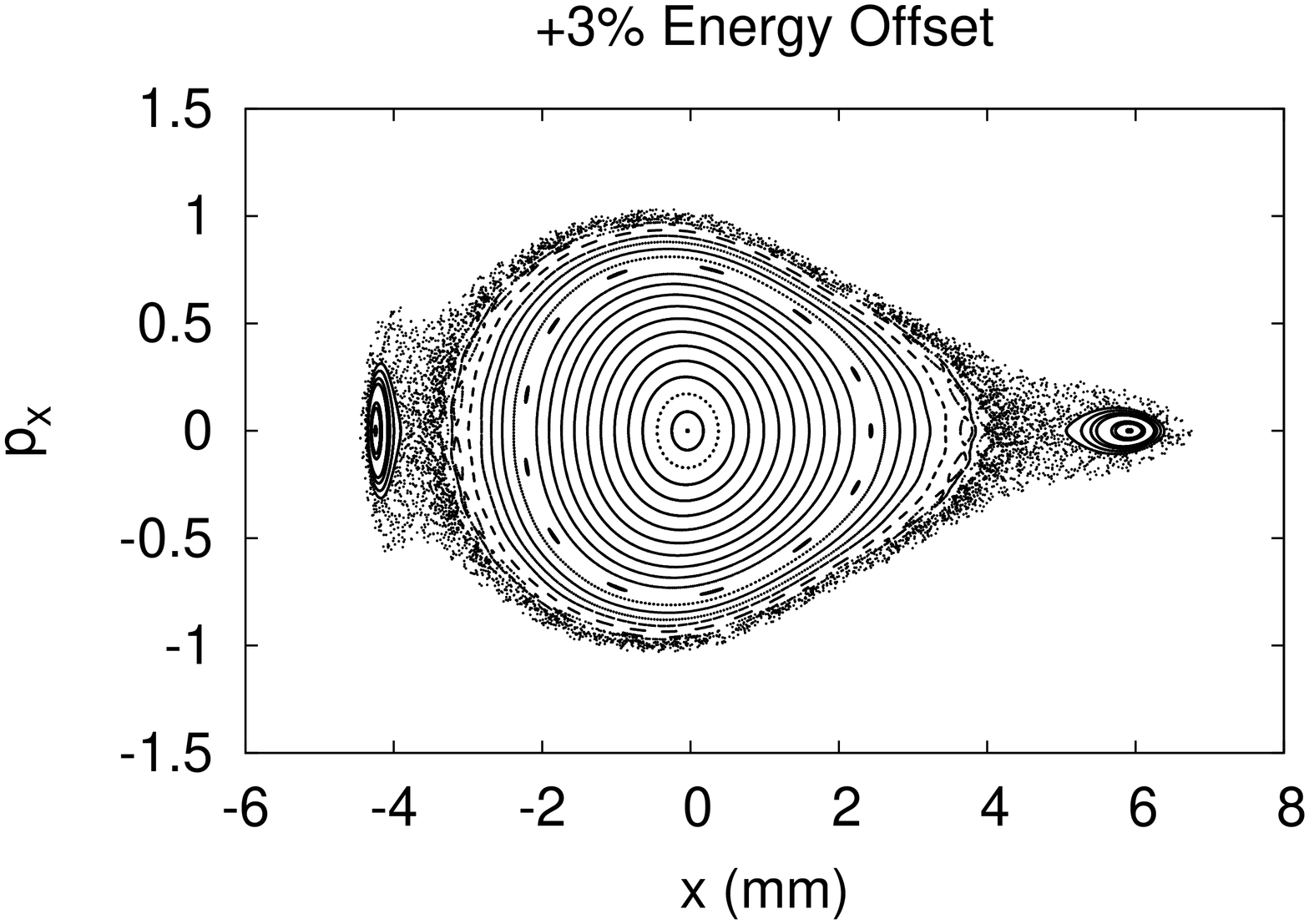}\\
\includegraphics[angle=-90,width=0.33\columnwidth]{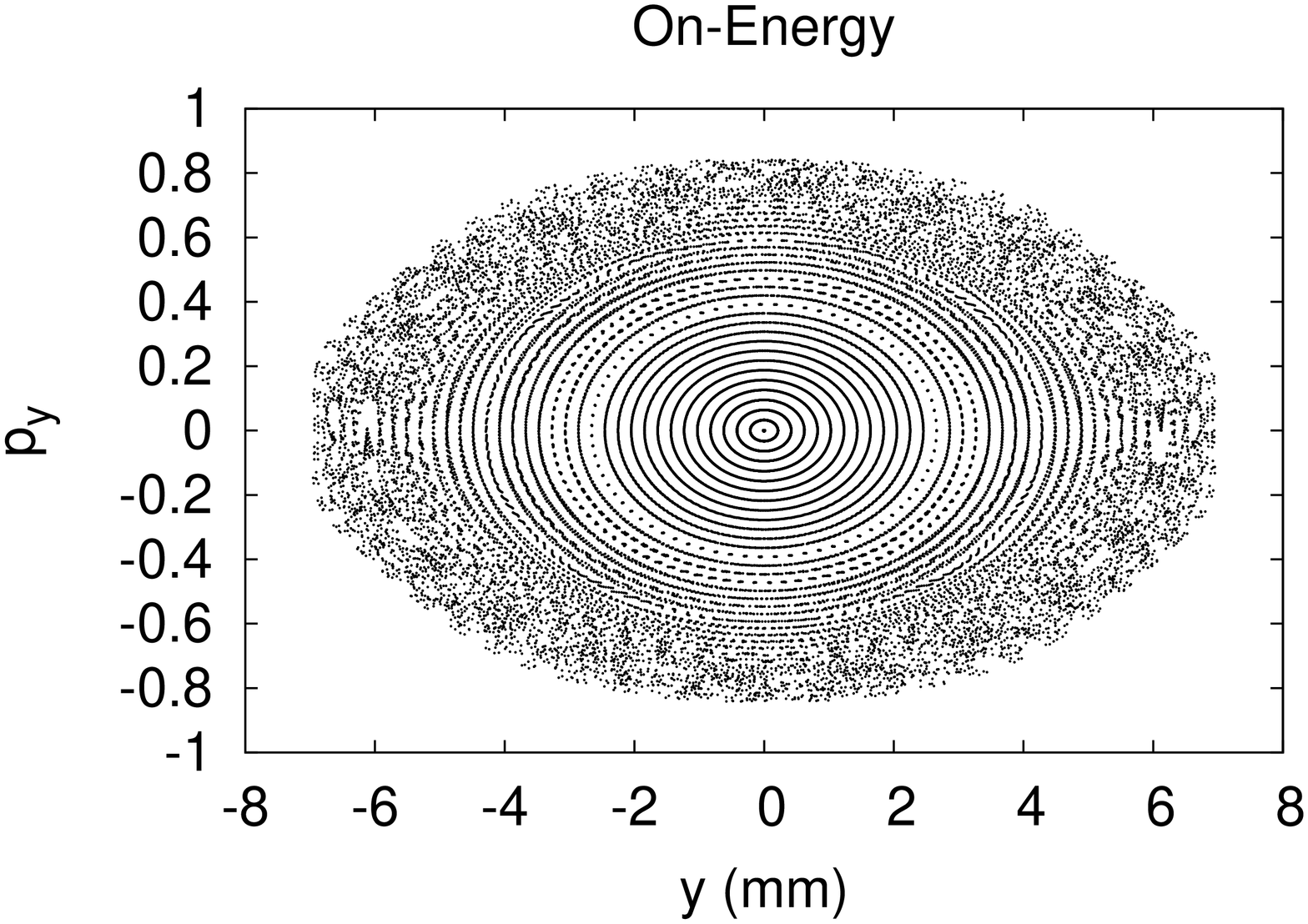}\hfill
\includegraphics[angle=-90,width=0.33\columnwidth]{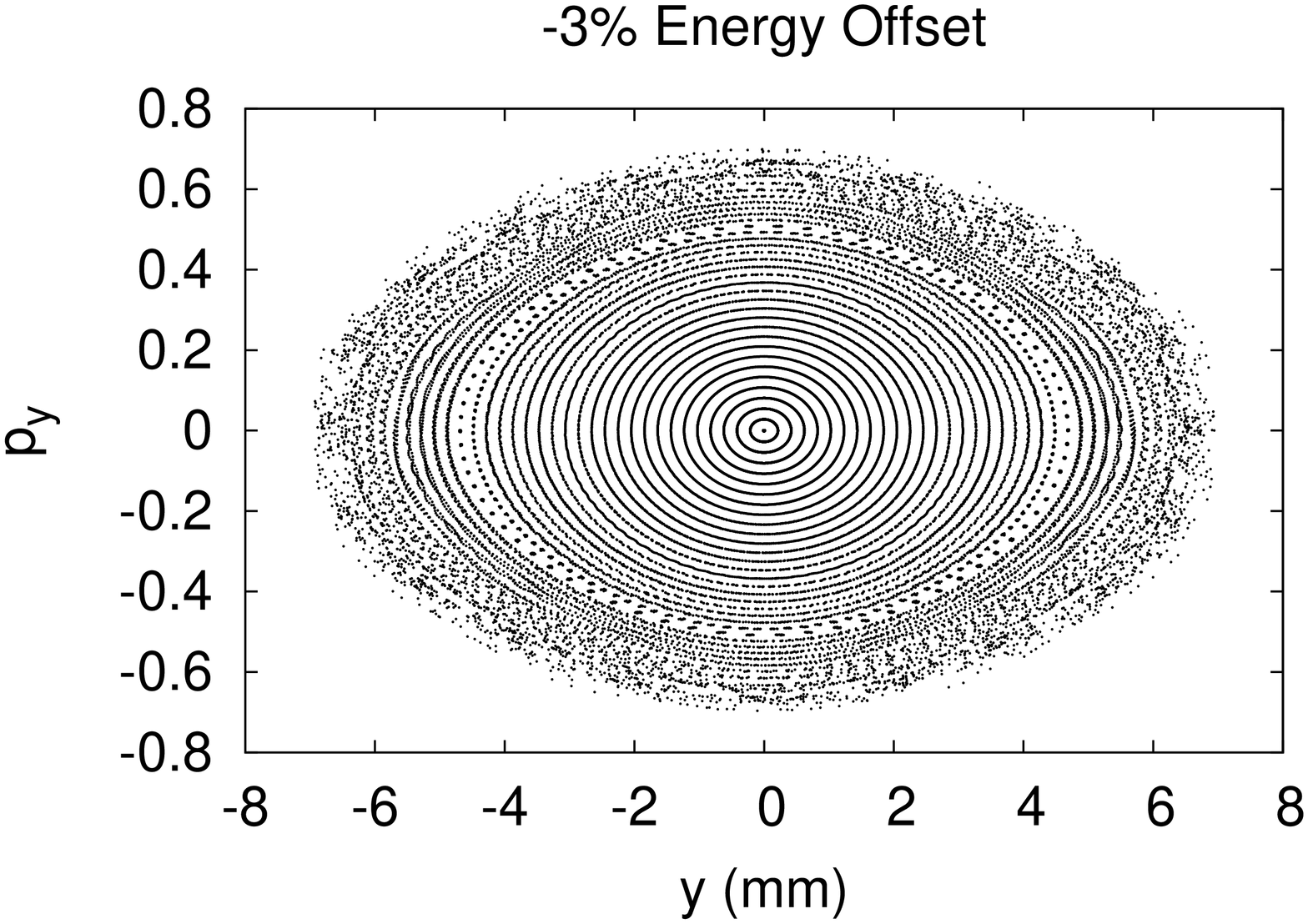}\hfill
\includegraphics[angle=-90,width=0.33\columnwidth]{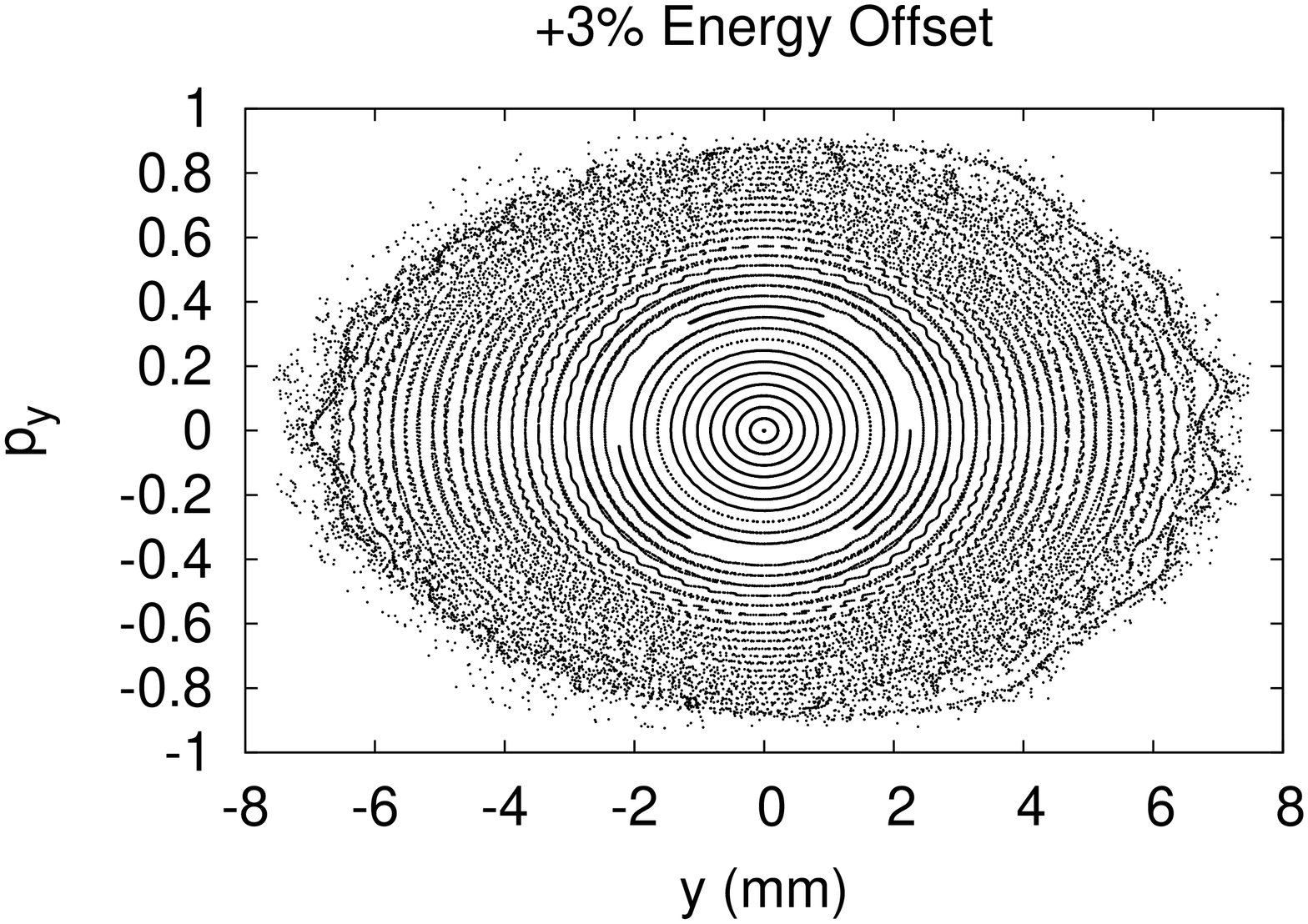}
\caption{Horizontal and vertical phase space portraits for SLS upgrade on-energy, and at $-3\%$ and $+3\%$
energy defect.
\label{fig:dc01a-ps}}
\end{figure}

Shown in Fig.~\ref{fig:dc01a-ps} are $x$-$p_x$ and $y$-$p_y$ phase space portraits for the optimized SLS upgrade lattice.  
The portraits are calculated using 4D tracking for $100$ turns.  
The lack of large resonance islands and lack of thick chaotic layers inside the stable region is a positive result that should
contribute to the robustness of the solution when misalignments are added.

\begin{figure}
\includegraphics[angle=0,width=\columnwidth]{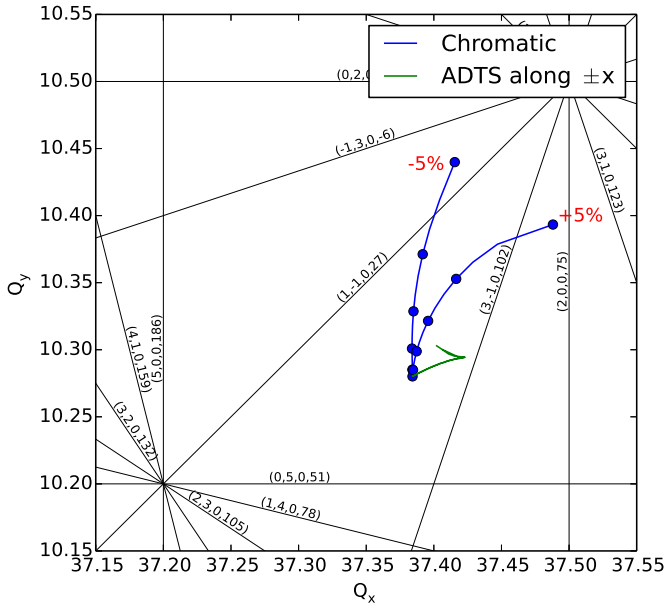}
\caption{SLS upgrade: Chromatic footprint from $-5\%$ to $+5\%$ and ADTS along $x$ from the $-x$ DA to the $+x$ DA. 
Blue points are $1\%$ increments.  
Resonance lines are labeled $\left(p,\,q,\,r,\,n\right)$ where $pQ_x+qQ_y+rQ_s=n$.
Low order resonance lines and higher order lines are plotted.  Higher order resonance lines excluded
by periodicity $3$ are not shown.
\label{fig:dc01a-fp}}
\end{figure}

Figure~\ref{fig:dc01a-fp} shows the chromatic
tune footprint and ADTS along $\pm x$.  The chromatic tunes are calculated by linearizing the
off-energy optics.  The ADTS is calculated by summing element-by-element phase advances in normal mode coordinates.
The ADTSs along $+x$ and $-x$ mostly overlap.

\subsection{CANDLE}
CANDLE is a proposed $3$ GeV, $216$ m Armenian light source project \cite{candle} providing $8.54$ nm horizontal beam
emittance.  According to the recent developments in storage ring lattice design and magnet technologies a new upgrade prototype 
has been designed \cite{candle-esls}, which is constructed of sixteen 4BA cells and provides $1.1$ nm horizontal beam emittance.  
The study shown here is on this new $1.1$ nm prototype.
Each cell is composed of combined function bends with both quadrupole and sextupole moments.
Some properties are shown in Tab.~\ref{tab:lats}.
Two features which contribute to CANDLE's nonlinearities are: 1) The sextupole moments are spread out across
a broad phase advance. 2) All sextupole moments are in dispersive regions.

\begin{figure}
\includegraphics[angle=-90,width=\columnwidth]{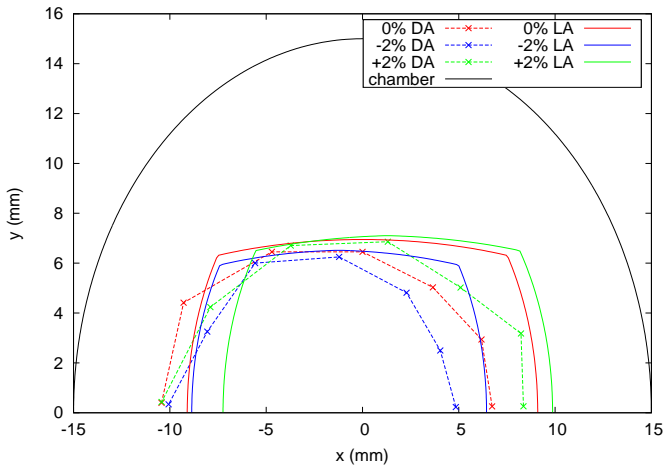}
\caption{Dynamic aperture (DA) and linear aperture (LA) for CANDLE at $0\%$, $-2\%$, and $2\%$ energy offset.
At the injection point $\beta_x=3.3$ m and $\beta_y=1.8$ m.
The maxima throughout the machine are $\beta_x=9.1$ m and $\beta_y=8.4$ m.
\label{fig:candle-da}}
\end{figure}

\begin{figure}
\includegraphics[angle=0,width=\columnwidth]{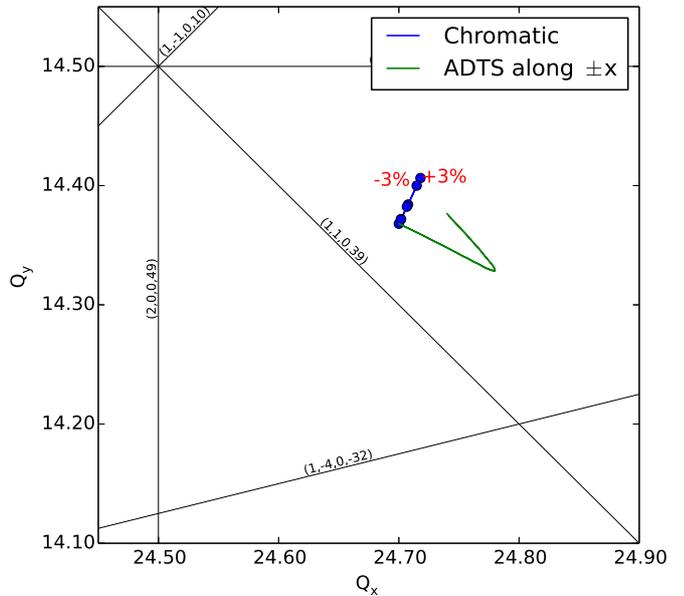}
\caption{CANDLE: Chromatic footprint from $-3\%$ to $+3\%$ and ADTS along $x$ from the $-x$ DA to the $+x$ DA.  ADTS
is calculated with a small vertical offset to allow for accurate calculation of vertical tune.
Higher order resonance lines excluded by periodicity $16$ are not shown.
\label{fig:candle-fp}}
\end{figure}

The dynamic aperture on-energy and at $\pm 2\%$ are shown in Fig.~\ref{fig:candle-da}.  The horizontal ADTS and
chromatic footprint out to $\pm3\%$ are shown in Fig.~\ref{fig:candle-fp}.

\begin{figure}
\includegraphics[angle=-90,width=\columnwidth]{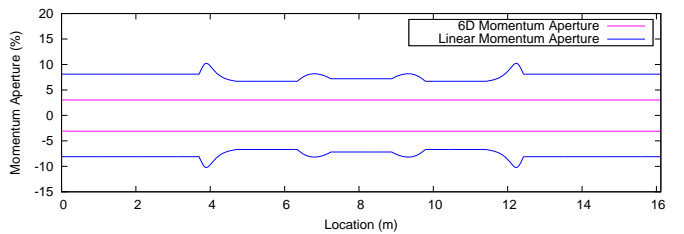}
\caption{Momentum aperture and linear momentum aperture for one period of the CANDLE lattice.  Linear momentum aperture is calculated
from Twiss optics linearized about the on-energy particle.  The RF voltage is set such that the linear calculation
of the RF bucket depth is $\pm3\%$.
\label{fig:candle-ma}}
\end{figure}

\begin{figure}
\includegraphics[angle=-90,width=0.33\columnwidth]{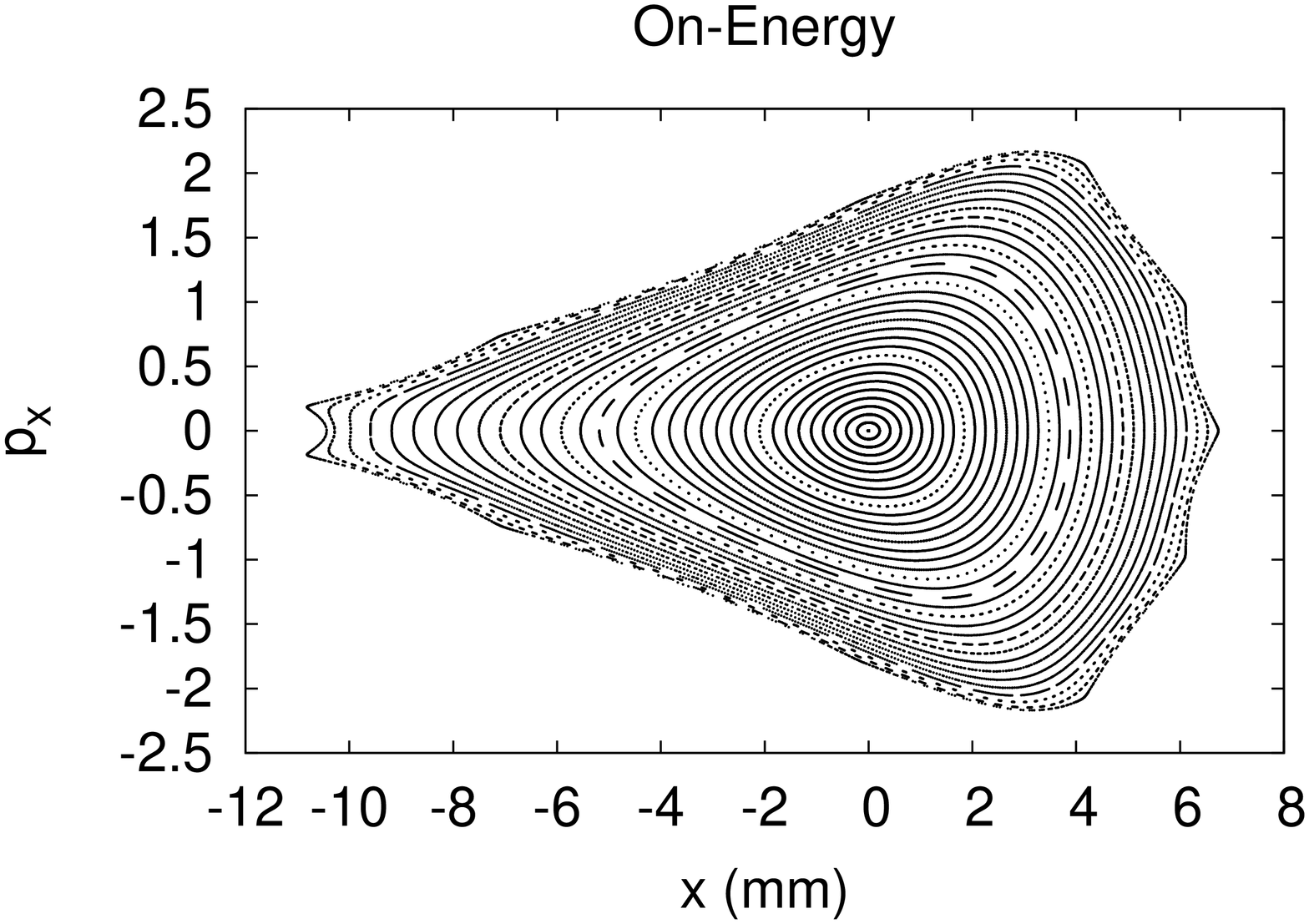}\hfill
\includegraphics[angle=-90,width=0.33\columnwidth]{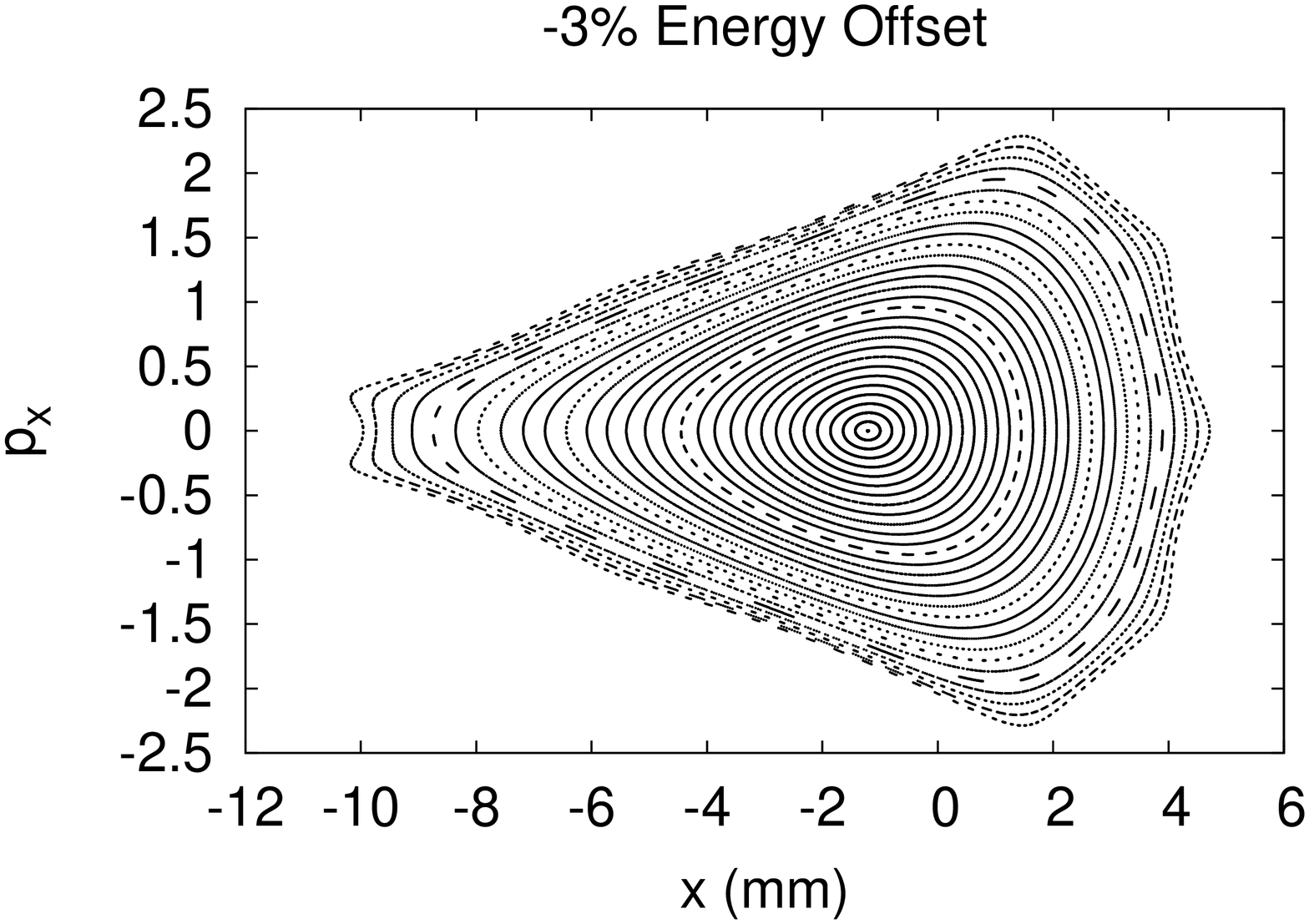}\hfill
\includegraphics[angle=-90,width=0.33\columnwidth]{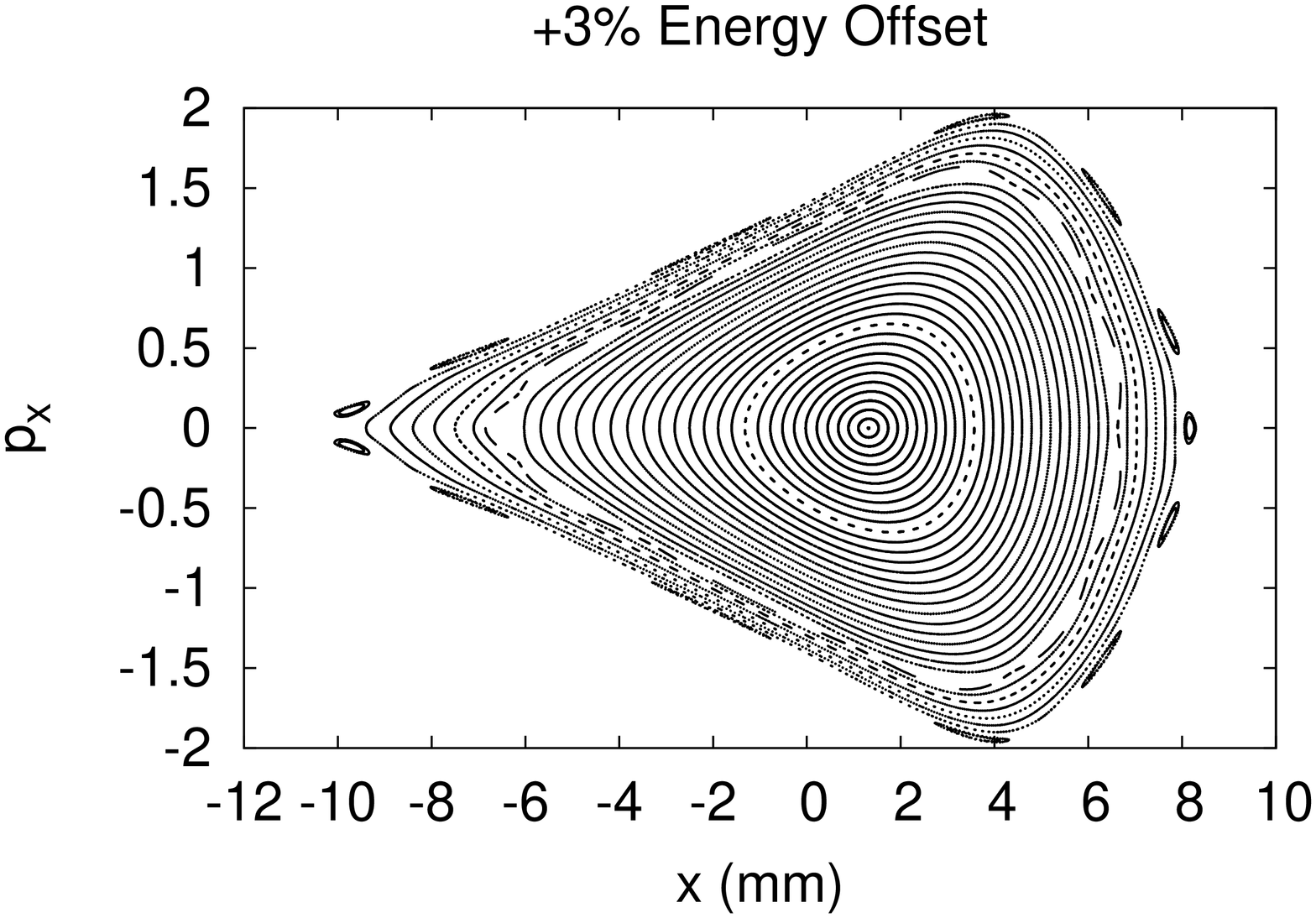}\\
\includegraphics[angle=-90,width=0.33\columnwidth]{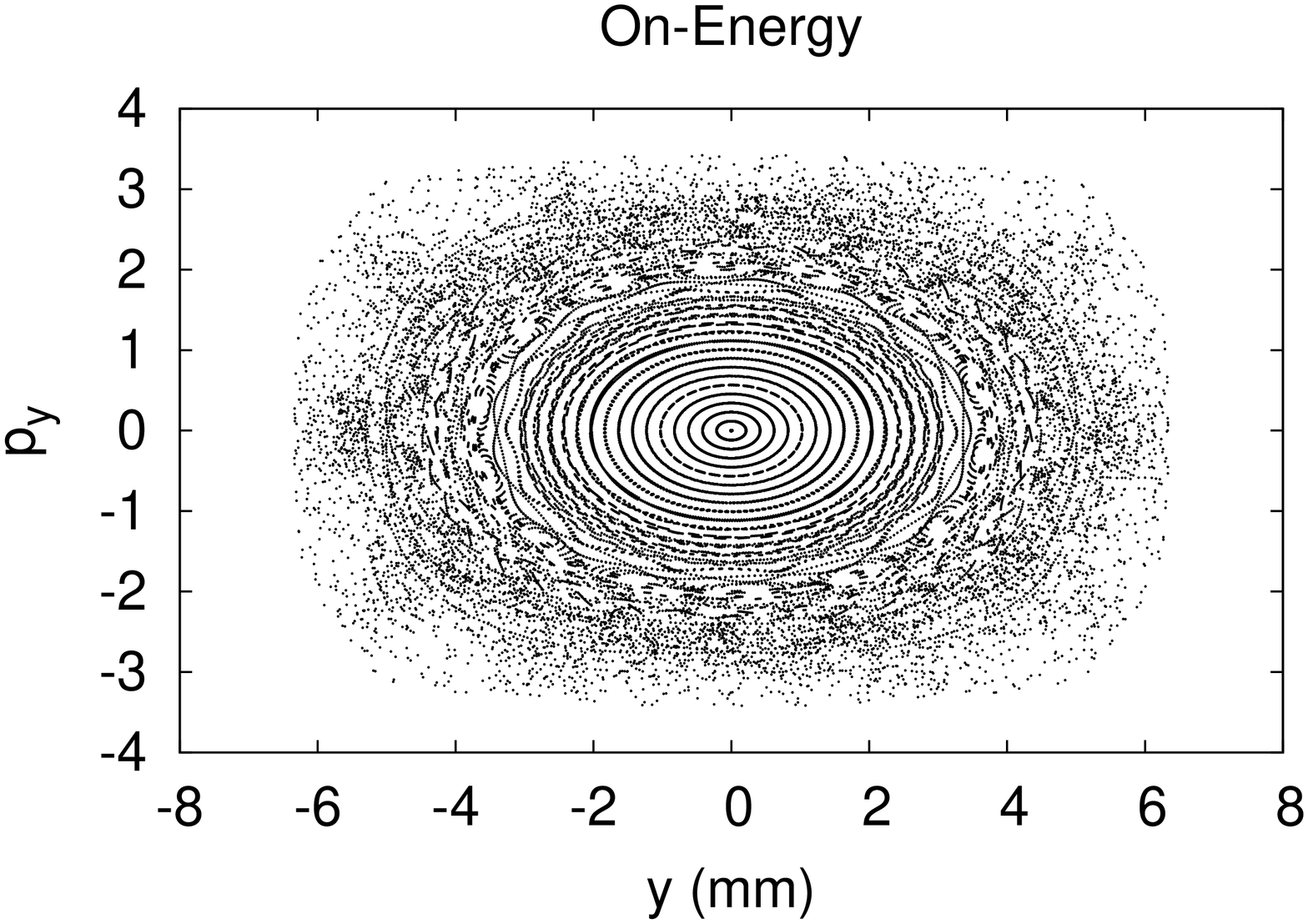}\hfill
\includegraphics[angle=-90,width=0.33\columnwidth]{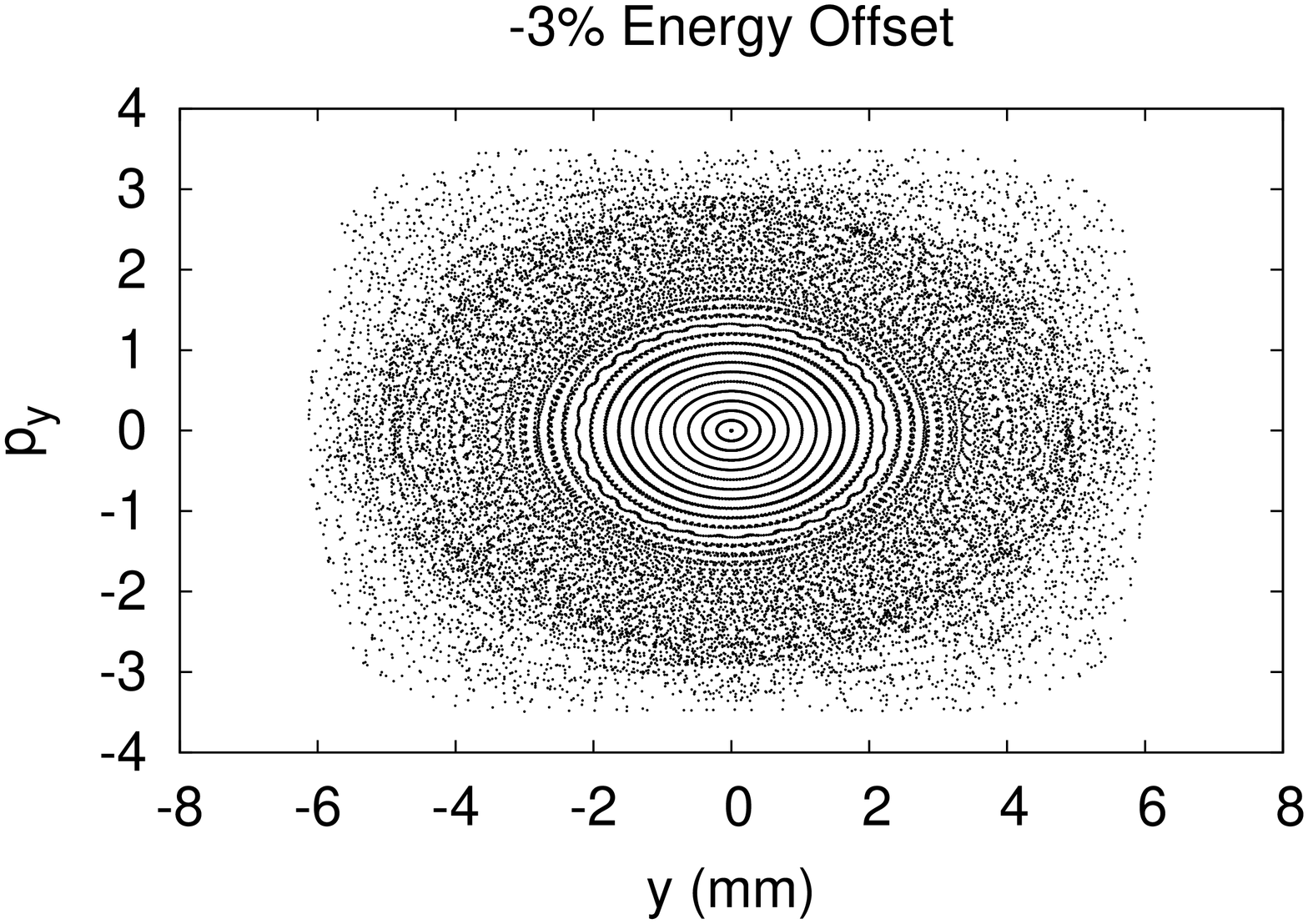}\hfill
\includegraphics[angle=-90,width=0.33\columnwidth]{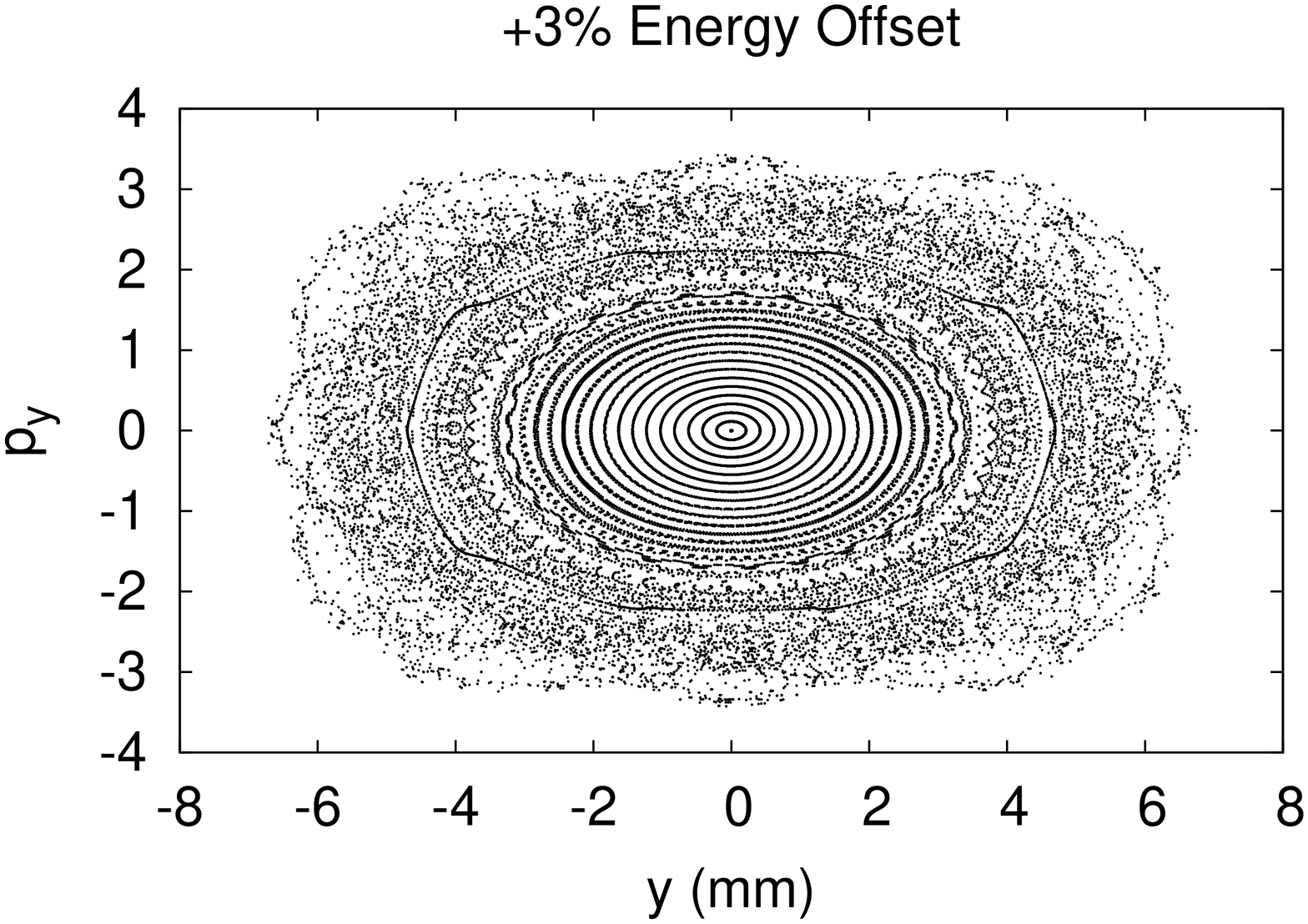}
\caption{Horizontal and vertical phase space portraits for CANDLE on-energy, and at $-2\%$ and $+2\%$
energy defect.
\label{fig:CANDLE-ps}}
\end{figure}

Touschek lifetime results and assumptions are shown in Tab.~\ref{tab:calc-tl}.
The momentum aperture calculated from 6D tracking is shown in Fig.~\ref{fig:candle-ma}.  The aperture
is determined entirely by the RF bucket and not limited by the transverse optics.
The phase space portraits are shown in Fig.~\ref{fig:CANDLE-ps}.
Optimizing the $\pm 2\%$ dynamic aperture and constraining
the chromatic tune footprint to $\pm 3\%$ has successfully optimized the global momentum aperture to at least  $\pm 3\%$.

\section{Tolerance to Machine Misalignments}\label{sec:ma}
The tolerance of the optimized SLS upgrade lattice to machine misalignments is tested.
The genetic algorithm is applied to the ideal lattice and a single solution is selected by the lattice
designer.  The lattice is then misaligned
according to Tab.~\ref{tab:ma} and corrected as described below.  The beam lifetime and
on-energy dynamic aperture of the resulting misaligned and corrected lattice is calculated.
This procedure is repeated for many misalignment seeds.
\begin{table}[h]
\begin{tabular*}{1.00\columnwidth}{@{\extracolsep{\fill}}lcc}
\hline
\hline
Property                          & Relative to    & Distribution $\sigma$ \\
\hline
Quad. \& sext. tilt               & girder         & $50$ $\mu$rad \\
Quad. \& sext. horiz. offset      & girder         & $20$ $\mu$m\\
Quad. \& sext. vert. offset       & girder         & $20$ $\mu$m\\
Bend \& anti-bend tilt            & girder         & $50$ $\mu$rad \\
Bend \& anti-bend horiz. offset   & girder         & $20$ $\mu$m\\
Bend \& anti-bend vert. offset    & girder         & $20$ $\mu$m\\
Girder tilt                       & lab            & $50$ $\mu$rad\\
Girder horiz. offset              & lab            & $50$ $\mu$m\\
Girder vert. offset               & lab            & $50$ $\mu$m\\
LGB tilt                          & lab            & $50$ $\mu$rad\\
LGB horiz. offset                 & lab            & $20$ $\mu$m\\
LGB vert. offset                  & lab            & $20$ $\mu$m\\
\hline
\hline
\end{tabular*}
\caption{Misalignments are drawn from a random Gaussian distribution, subject to
a $2$-$\sigma$ cutoff.  Misalignments which exceed the cutoff are re-drawn.
\label{tab:ma}}
\end{table}

The correction procedure begins by flattening the horizontal and vertical orbits using an
orbit response matrix and SVD.  Then a simultaneous horizontal phase, vertical phase, and
horizontal dispersion correction is applied using a combined phase and dispersion response matrix.
The residual coupling after these corrections ranges from $0.4$ to $3.2\%$.  A dedicated
coupling correction is not included in this study.

\begin{figure}
\includegraphics[angle=-90,width=\columnwidth]{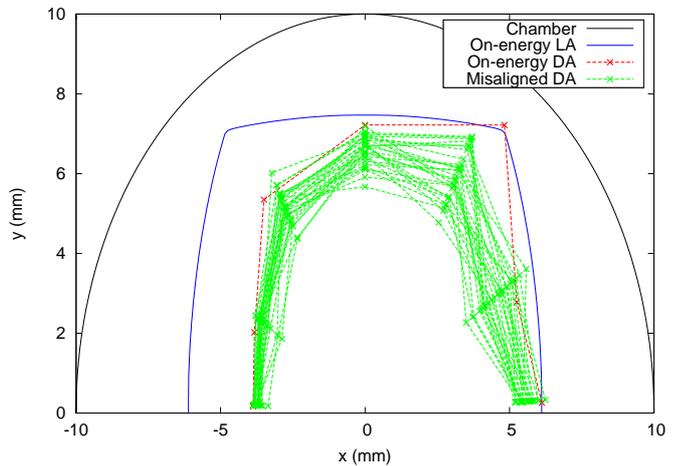}
\caption{On-energy dynamic aperture for ideal lattice and $30$ misaligned and corrected lattices.
At the injection point $\beta_x=3.3$ m and $\beta_y=6.5$ m.
The maxima throughout the machine are $\beta_x=8.7$ m and $\beta_y=11.7$ m.
\label{fig:mada}}
\end{figure}

\begin{figure}
\includegraphics[angle=-90,width=\columnwidth]{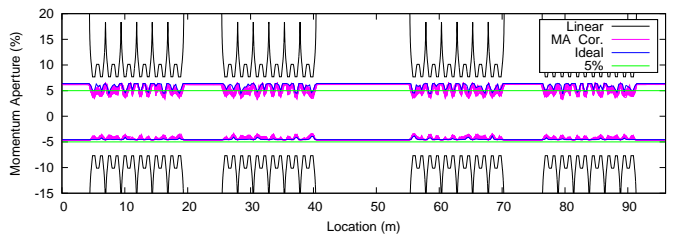}
\caption{Momentum aperture from 6D tracking for ideal lattice and $30$ misaligned and corrected lattices.
\label{fig:mama}}
\end{figure}

So $30$ misalignment seeds are generated and corrected.  
One of these seeds fails to have a closed orbit and is discarded.
The on-energy dynamic aperture and momentum aperture for the remaining $29$ are calculated and the results
are shown in Fig.~\ref{fig:mada} and Fig.~\ref{fig:mama}.
$50\%$ of the misaligned and corrected lattices
have a lifetime longer than $3.8$ hr, and $95\%$ have a lifetime longer than $3.6$ hr.  The reference lifetime
is $4.4$ hr. 

The misalignment and correction procedure applied here is pessimistic.  The fully developed
SLS upgrade misalignment model will take into account
that the LGBs will be aligned relative to the girders, and the girders relative
to one another.
Furthermore, coupling correction and vertical dispersion correction
will be applied in the actual machine.  Despite the pessimistic scenario, the calculated dynamic aperture
and lifetimes of the misaligned and corrected lattices are acceptable.  From this we conclude that the 
sextupole and octupole scheme generated by the genetic algorithm is sufficiently robust against imperfections.

\begin{figure}
\includegraphics[angle=-90,width=\columnwidth]{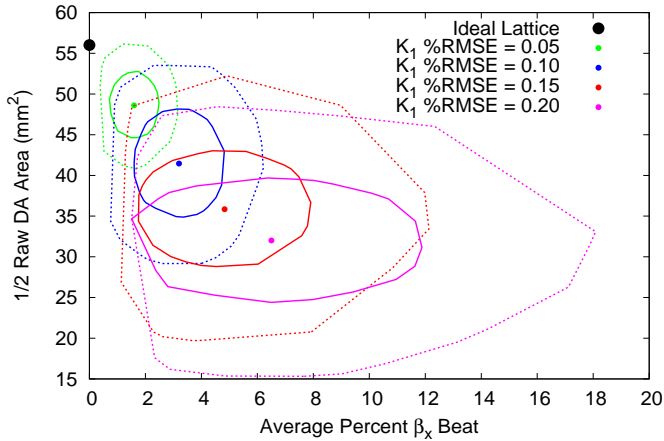}
\caption{Area of the $+y$ dynamic aperture versus percent beta beating induced by quadrupole gradient errors.
The solid enclosure is the convex hull containing $50\%$ of the seeds nearest the mean,
the dashed enclosure contains $90\%$ of the seeds.
\label{fig:beta_beating}}
\end{figure}

The sensitivity of the chromaticity correction scheme to beta beating is tested by applying gradient
errors to the quadrupole moments in quadrupoles and gradient bends.  
Errors with RMS values of $0.05\%$, $0.10\%$, $0.15\%$, and $0.20\%$, subject
to a $2$-$\sigma$ cutoff, are tested.  No corrections are applied.
$1000$ seeds are generated for each of the four cases.  For each seed, the on-energy dynamic aperture
and mean percent beta beat is calculated.  Plotted in Fig.~\ref{fig:beta_beating} is the
mean for each case and the convex hulls that contain $50\%$ and $90\%$ of the seeds closest to the mean.
In the original SLS, beta-beating is measured to be $2\%$ \cite{beta-beat:aiba}.  We therefore anticipate a reduction in the 
on-energy dynamic aperture area in the SLS upgrade due to beta-beating of less than $20\%$.

\section{Conclusion}
The genetic algorithm presented here offers a robust and computationally affordable
technique for generating globally optimal chromaticity correction schemes for diffraction limited light sources.
The resulting correction schemes have good on-energy dynamic aperture which should help injection efficiency
and give a wide momentum aperture for long beam lifetime.  The schemes are sufficiently robust against misalignments.

One feature of this algorithm is the use of dominance constraints to encourage individuals in the early population
to take on properties that will later on contribute to healthy objective values. 

A second feature is the use of off-energy dynamic aperture along with tune footprint constraints as a proxy
for the computationally expensive direct momentum aperture calculation.

Based on development efforts at SLS and results shared by the CANDLE collaboration \cite{candle-esls},
this genetic algorithm delivers results that are as good or better than those obtained by applying 2nd order 
resonant driving term minimization.  The genetic algorithm
converges in a couple days on commonly available computing resources.  This turn-around time is comparable to that 
required for a lattice designer to develop a scheme using resonant driving term minimization.  
Thus the genetic algorithm presented here is a practical solution for optimizing sextupole and octupole strengths
in a diffraction limited light source project.

\begin{acknowledgments}
I would like to thank David Sagan of Cornell University for supporting the development of this 
genetic algorithm through his work on the {\tt Bmad} \cite{bmad:2006} 
library.  Many aspects of this work benefited greatly from discussions with Masamitsu Aiba and Michael B\"{o}ge.
I would like to thank Andreas Streun for his mentorship in storage ring nonlinearities, and for the well-designed SLS upgrade lattice.
I would like to thank Artsrun Sargsyan for the opportunity to contribute to the development of CANDLE.
\end{acknowledgments}

\bibliography{prstab-moga.bib}

\end{document}